\documentclass{iopjournal}

\usepackage{graphicx} 
\usepackage{amsmath, physics, amsfonts, amssymb, mathtools, bm}
\usepackage{hyperref} 
\graphicspath{{./figures/}}

\newcommand{\N}{\mathbb{N}}

\newcommand{\Exp}{\mathbb{E}}
\newcommand{\Prb}{\mathbb{P}}

\newcommand{\Norm}{\mathcal{N}}
\newcommand{\Orb}{\mathcal{O}}
\newcommand{\e}{\mathrm{e}}

\newcommand{\ve}{\varepsilon}
\newcommand{\qqquad}{\qquad\quad}
\newcommand{\qqqquad}{\qqquad\quad}
\newcommand{\qqqqquad}{\qqqquad\quad}

\newcommand{\epsilonhat}{\hat{\epsilon}}

\DeclareMathOperator{\Pois}{Pois}

\usepackage{pifont}
\newcommand{\cmark}{\ding{51}}
\newcommand{\xmark}{\ding{55}}

\usepackage[table,dvipsnames]{xcolor}
\definecolor{r}{RGB}{255,182,193}
\definecolor{g}{RGB}{152,251,152}
\definecolor{o}{RGB}{255,222,173}
\definecolor{w}{RGB}{255,255,224}
\definecolor{b}{RGB}{173,216,230}

\usepackage{graphbox, scalerel, enumitem, multirow}

\usepackage{epigraph}
\usepackage{nicematrix}
\newcommand{\y}{\Block[fill=green!10]{}{\cmark}}
\newcommand{\n}{\Block[fill=red!10]{}{\xmark}}

\newcommand{\vertiii}[1]{{\left\vert\kern-0.25ex\left\vert\kern-0.25ex\left\vert #1 
    \right\vert\kern-0.25ex\right\vert\kern-0.25ex\right\vert}}

\usepackage[sorting=none,style=numeric-comp]{biblatex} 
\renewbibmacro{in:}{%
  \ifboolexpr{%
     test {\ifentrytype{article}}%
     or
     test {\ifentrytype{inproceedings}}%
  }{}{\printtext{\bibstring{in}\intitlepunct}}%
}

\begin{document}

\articletype{Paper} 

\title{Quasi-stationary and quasi-ergodic distributions in the Pelikan random map}

\author{Samuel Brevitt$^{1,*}$\orcid{0000-0002-9257-6335} and Rainer Klages$^{1,2}$\orcid{0000-0003-3811-3070}}

\affil{$^1$Centre for Complex Systems, School of Mathematical Sciences, Queen Mary University of London, Mile End Road, London E1 4NS, United Kingdom}

\affil{$^2$London Mathematical Laboratory, 8 Margravine Gardens, London W6 8RH, United Kingdom}

\affil{$^*$Author to whom any correspondence should be addressed.}

\email{s.brevitt@qmul.ac.uk}


\begin{abstract}
In this paper we present a concrete example of a substochastic discrete-time Markov chain on a countable state space producing a spectrum of infinitely many quasi-stationary distributions (QSDs) for generic parameter values, with each QSD supporting a distinct escape rate. Our system is motivated by an open variant of the Pelikan dynamical system, a random map introduced in the 1980s. These QSDs, and their stability to perturbative random noise, are tested in numerical simulations. The existence of unique QEDs is also established for some parameter values.
\end{abstract}


\section{Introduction}

\subsection{Motivation} \label{sec:intro}

Given a discrete-time Markov chain, there are several
objects and measurements
which are of interest to us. Principal among these is the invariant measure, a distribution over the states of the chain which is invariant under the action of the stochastic matrix governing
the chain. Formally, if a Markov chain $(X_n)_{n=0}^\infty$ on a state space $\Omega$ is defined by a stochastic (henceforth `closed') matrix $M$, such that $M_{ji} := \Prb[X_{n+1}=j\,|\,X_n=i]$, an invariant measure is an eigenvector $\mu^*$ of $M$ with $\norm{\mu^*}_1 = 1$ satisfying
\begin{equation}
    M\mu^*=\mu^*.
\end{equation}
A consequence is that if the chain is ergodic (if finite: irreducible and aperiodic), then the invariant measure is also the \emph{limiting measure},
\begin{equation}
    \lim_{n\to\infty} M^n\mu = \mu^* \quad \textrm{for any initial positive $\mu$ with $\norm{\mu}_1 = 1$,}
\end{equation}
and the \emph{time-average} of $X_n$,
\begin{equation}
    \lim_{N\to\infty} \frac{1}{N} \#\{ X_n = i\,|\, n\in\{1, \dots, N\} \} = \mu^*_i \quad \textrm{(a.s.)}
\end{equation}
regardless of the initial state $X_0$ \cite{norris_markov_1997, meyer_matrix_2023}.

If the matrix $M$ is substochastic (or `open'), on the other hand, some probability mass is lost at each timestep (in such circumstances we consider some realisations of the process to have been `killed' and $X_n\notin\Omega$), and therefore we have no invariant measure. Assuming $M$ is irreducible, aperiodic, and finite, we rather have a \emph{conditionally invariant measure}, or \emph{quasi-stationary distribution} (QSD) \cite{collet_quasi-stationary_2013, castro_existence_2024, darroch_quasi-stationary_1965}, satisfying
\begin{equation}
    M\mu^* = \lambda\mu^*, \quad 0<\lambda<1,
\end{equation}
for $\lambda$ the largest eigenvalue of $M$, which the Perron-Frobenius theorem guarantees is real, positive, simple and unique. Under these conditions, the eigenvector $\mu^*$ (the QSD) is positive, and is the only positive eigenvector \cite{meyer_matrix_2023}.
The QSD is also the conditional limiting measure, in the sense of the Yaglom limit
\begin{equation} \label{eq:yaglom}
    \lim_{n\to\infty} \frac{M^n\mu}{\norm{M^n\mu}_1} = \mu^* \quad
    \textrm{for any initial positive $\mu$,}
\end{equation}
but is generically \emph{not} the conditional time-average,
\begin{equation} \label{eq:qed-def}
    \lim_{N\to\infty} \Exp\left[ \frac{1}{N} \#\{ X_n = i\,|\, n\in\{1, \dots, N\} \} \;\middle|\; X_N\in\Omega \right] = \bm{q}_i,
\end{equation}
which is instead given by the element-wise product $\bm{q}_i=b_i\mu^*_i$ of $\mu^*$ with the left-eigenvector of the matrix, satisfying
\begin{equation}
    M^Tb = \lambda b
\end{equation}
for the same $\lambda$ \cite{darroch_quasi-stationary_1965}. This product $\bm{q}$ is named the \emph{quasi-ergodic distribution} (QED) \cite{castro_existence_2024, colonius_quasi-ergodic_2021}. The probability of survival until time $n$ generically decays asymptotically exponentially with $n$, with
\begin{equation}
    \Prb[X_n\in\Omega] = \norm{M^n\mu}_1 \sim \lambda^n = \e^{-\gamma_{\textrm{esc}}n}, \quad \textrm{for} \quad \gamma_\textrm{esc}:=-\ln\lambda,
\end{equation}
asymptotically as $n\to\infty$.


When $M$ is not finite, however, the existence and uniqueness of $\lambda$, $\mu^*$, and $\bm{q}$ is not guaranteed: the operator $M$ is not necessarily compact, and therefore its spectrum is generically not simply a discrete set of eigenvalues. Some results along these lines were identified for countable matrices in \cite{seneta_quasi-stationary_1966, vere-jones_ergodic_1967}, although these results remain much less well known than those for finite matrices.

From \cite{seneta_quasi-stationary_1966}, we obtain the following: QSDs are still, by definition, given by eigenvectors of the conditional transfer operator; however, these need not be unique: there may be multiple positive eigenvectors, with distinct corresponding positive eigenvalues. At best, if $M$ is $R$-positive, in the formalism of Vere-Jones \cite{vere-jones_ergodic_1967}, with convergence parameter $R$, then the Yaglom limit \eqref{eq:yaglom} converges from any initial \emph{fixed state} $X_0=i$ (Thm.~3.1 in \cite{seneta_quasi-stationary_1966}) to a unique QSD whose eigenvalue
$\lambda = 1/R$;
but from an arbitrary initial \emph{distribution}, all bets are off, and we only have that other QSDs, if they exist, have eigenvector $\lambda \in [1/R,1)$, and are converged to by some set of initial conditions (Thm.~4.1 in \cite{seneta_quasi-stationary_1966}). Similarly, if $M$ is $R$-positive, we have existence and uniqueness of a QED given by $\bm{q}_i=b_i\mu^*_i$ \cite{seneta_quasi-stationary_1966}.



\subsection{Outline of results}

In this paper we identify a simple discrete-time countable substochastic Markov chain which displays, for generic parameter values, a continuous spectrum of QSDs, each supported by its own eigenvalue $\lambda$, corresponding to a distinct rate of escape from the system. We introduce our system in Sec.~\ref{sec:sec1}. Our system takes the form of a biased random walk on the half-line, with a form of stochastic resetting and escape. This example is motivated by a classical spatially-continuous random dynamical system, the Pelikan map, proposed as a simple example in \cite{pelikan_invariant_1984}, whose importance was highlighted in \cite{sato_anomalous_2019,yan_transition_2024}. Thanks to recent results \cite{sato_anomalous_2019, yan_transition_2024, monthus_explicit_2025, brevitt_weak_2026}, it is known that the Pelikan map transitions between modes of strong chaos (in the language of Markov chains: positive recurrence) and a form of intermittency called `on-off' intermittency \cite{pikovsky_interaction_1984, fujisaka_new_1985, fujisaka_stability_1986, pikovsky_symmetry_1991, ott_blowout_1994, platt_-off_1993, heagy_characterisation_1994, hata_exactly_1997, goverse_intermittent_2026}, corresponding in a stochastic sense to Markov null-recurrence. Accordingly, we demonstrate in Sec.~\ref{sec:F00-calculations} that our open system transitions between modes of positive $R$-recurrence and $R$-transience,
under variation of its two parameters: the random walk bias $p$, and the hole size $\ve$.
Importantly, the multiplicity of QSDs occurs regardless of which of these modes the system is in. In Sec.~\ref{sec:2} we analytically calculate the QSDs of the system, where we also verify their invariance in simulations in Sec.~\ref{sec:qsd-simulations}. In Sec.~\ref{sec:shape} we consider the asymptotic shape of these QSDs, and identify it as an invariant property under the system dynamics. We assess the stability of our QSDs to perturbation by random noise in Sec.~\ref{sec:stability}. We also explicitly calculate, and demonstrate the existence and uniqueness of, QEDs in the region where $M$ is $R$-positive, in Sec.~\ref{sec:qed}.


\section{Pelikan system} \label{sec:sec1}

The system we consider in this article is supported on the countably infinite discrete state space, indexed by $k\in\N\cup\{0\}$, with transfer operator
\begin{equation} \label{eq:pelikan-markov}
    M_{ji} = \Prb[X_{n+1}=j\,|\,X_n=i] := p\,\bm{1}_{j=i-1} + q\,\bm{1}_{j=i+1} + 2^{-j-1}p(1-\ve)\,\bm{1}_{i=0},
\end{equation}
depending on two parameters $p$ and $\ve$, and with $q:=1-p$. This can be seen as the union of two basic systems: one, $M_{\textrm{exp}}$, given by
\begin{equation} \label{eq:exp}
    (M_{\textrm{exp}})_{ji} = \begin{cases}
        \bm{1}_{j=i-1}, & i\neq 0, \\
        (1-\ve)\,2^{-j-1}, & i=0,
    \end{cases}
\end{equation}
which moves state $i$ to state $i-1$, unless $i=0$, in which case the probability mass is redistributed across all states in a geometric distribution,
with a proportion $\ve$ killed; and another, $M_{\textrm{cont}}$, given by
\begin{equation}
    (M_{\textrm{cont}})_{ji} = \bm{1}_{j=i+1}
\end{equation}
which always moves state $i$ to state $i+1$. The final matrix $M$ is given by $M = pM_\textrm{exp}+qM_\textrm{cont}$, which therefore dynamically represents a biased random walk on the half-line, with a form of sub-stochastic resetting at the state $0$.

\begin{figure}[ht]
    \centering
    \includegraphics[width=0.3\linewidth]{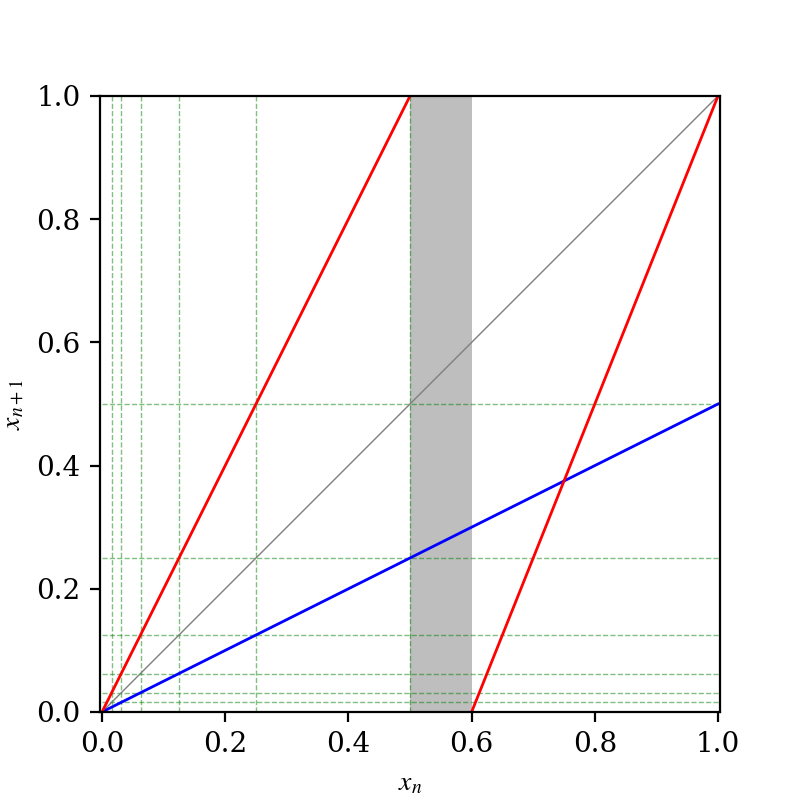}
    \includegraphics[width=0.69\linewidth]{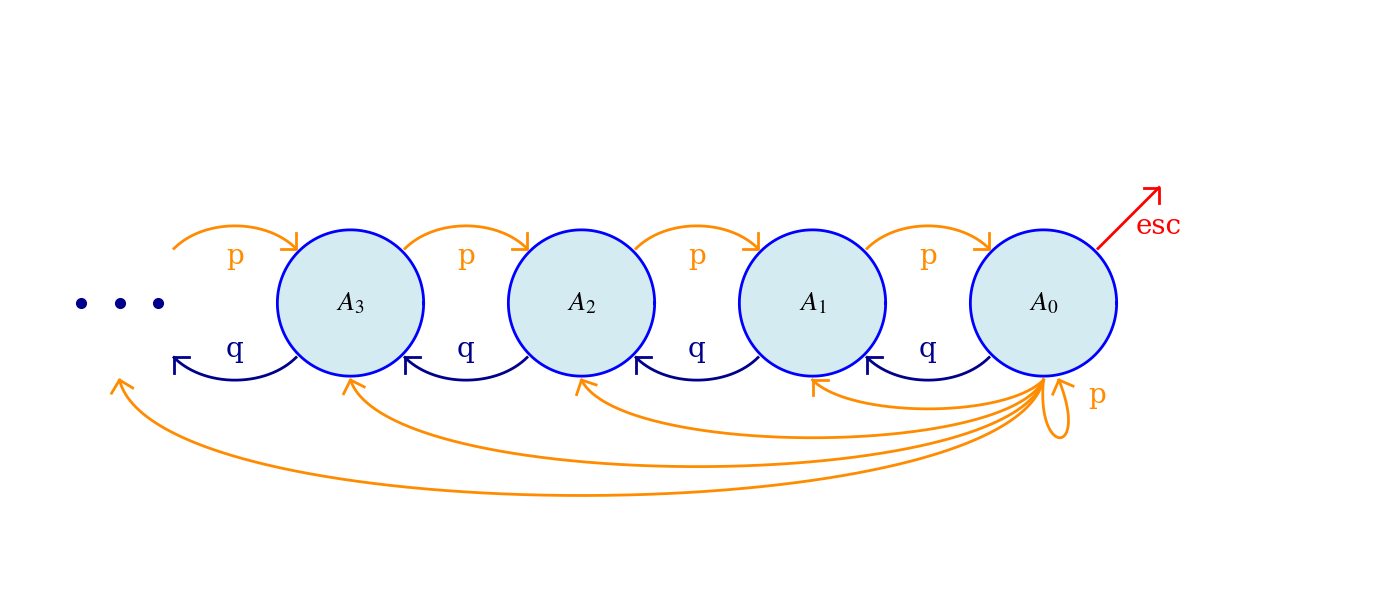}
    \caption{Left: The `open' Pelikan map \eqref{eq:pelikan-det}, with expanding and contracting branches shown in red and blue, respectively. The Markov partition \eqref{eq:partition} is shown in thin green dashed lines. The escape region is indicated in grey. Right: A schematic diagram of the Markov chain \eqref{eq:pelikan-markov}, which is the object of this paper's study.}
    \label{fig:fig1}
\end{figure}

This Markov process is motivated by the random, spatially continuous, dynamical system
\begin{equation} \label{eq:pelikan-det}
    T(x) = \left\{ \begin{matrix}
        \left\{ \begin{matrix}
            2x, & 0 \leq x \leq \frac12, \\
            \frac{x-(1+\ve)/2}{1-(1+\ve)/2}, & \frac{1+\ve}{2} \leq x \leq 1,
        \end{matrix} \right\} & \textrm{with probability $p$,} \\
        x/2, & \textrm{with probability $q$},
    \end{matrix} \right\}, \quad 0\leq x \leq 1.
\end{equation}
Random dynamical systems have long received attention within the study of dynamical systems, usually in the context of small random perturbations of otherwise interesting, and in particular chaotic, deterministic systems \cite{kifer_ergodic_1986, kifer_random_1988, kifer_small_1974, keller_stochastic_1982, kapitaniak_chaos_1990, baladi_spectra_1993, lasota_chaos_1994, arnold_random_1998, collet_enhancement_1999, klages_transitions_2002, freidlin_random_2012}; however, they may also be applied in a variety of abstract settings, such as for example in the study of iterated function systems, with geometric applications to the generation of fractals \cite{barnsley_fractals_1993}.
In the case $\ve=0$, the map \eqref{eq:pelikan-det} reduces to
\begin{equation} \label{eq:pelikan-closed}
    T_{\ve=0}(x) = \begin{cases}
        2x \ \textrm{(mod $1$)}, & \textrm{with probability $p$}, \\
        \quad x/2, & \textrm{with probability $q$},
    \end{cases}
\end{equation}
which is known as the Pelikan map, the properties of which were studied in \cite{pelikan_invariant_1984, sato_anomalous_2019, yan_transition_2024, monthus_explicit_2025}. Both the classical `closed' Pelikan map and our `open' generalisation \eqref{eq:pelikan-det} admit a Markov partition
\begin{equation} \label{eq:partition}
    \{A_k\} = \{ (2^{-k-1},2^{-k}) \,:\, k\in\N\cup\{0\} \}
\end{equation}
according to which the dynamics of the random map reduce to the countable Markov process \eqref{eq:pelikan-markov} \cite{brevitt_weak_2026}. The generalised `open' Pelikan map \eqref{eq:pelikan-det} and the Markov chain \eqref{eq:pelikan-markov} are shown in Fig.~\ref{fig:fig1}.

\subsection{Case \texorpdfstring{$\ve=0$}{ε=0}}

Thanks to results in \cite{pelikan_invariant_1984, yan_transition_2024}, we know the following information for the `closed' Pelikan map \eqref{eq:pelikan-closed}: for $p>\frac12$ the invariant density of the map, $\mu^*$,
is known, finite, and piecewise constant on the partition \eqref{eq:partition}, and was calculated to be
\begin{equation} \label{eq:invmeas}
    \mu^*_i = \mu^*(A_i)= \frac{2p-1}{3p-2} \left[ 2^{-i-1} - \left( \frac{1-p}{p} \right)^{i+1} \right]
\end{equation}
in \cite{yan_transition_2024}, which can be easily verified as being normalised.
The Lyapunov exponent of the system is also known and is positive in this regime, with $\lambda = (2p-1)\ln2$.
For $p=\frac12$ the map preserves an infinite invariant density -- i.e., an invariant density which is not normalisable \cite{aaronson_introduction_1997} -- and
the Lyapunov exponent
is zero. For $p<\frac12$ the only invariant density is the delta function at $x=0$, and the Lyapunov exponent is negative \cite{yan_transition_2024}.

The Markov representation of the closed Pelikan map \eqref{eq:pelikan-closed} is obtained by setting $\ve=0$ in \eqref{eq:pelikan-markov}, and is given by
\begin{equation} \label{eq:pelikan-closed-markov}
    M_{ji}^\textrm{$\ve=0$} = p\,\bm{1}_{j=i-1} + q\,\bm{1}_{j=i+1} + 2^{-j-1}p\,\bm{1}_{i=0},
\end{equation}
and has the following properties:
for $p>\frac12$, this Markov chain is positive recurrent, meaning all areas of the state space are visited a.s.\ infinitely often with finite expected recurrence time; for $p=\frac12$ it is null-recurrent, with all states visited a.s.\ infinitely often with \emph{infinite} expected recurrence time; and for $p<\frac12$ it is a transient Markov process, with all states experiencing an a.s.\ finite number of visits by the process as
$X_n\to\infty$.
The dynamical and stochastic properties of the closed system are thus summarised in Table~\ref{tab:table1}.

\begin{table}[ht]
    \centering
    \small
    \begin{NiceTabular}[hvlines]{cccccc}
        \Block{}{$\begin{matrix} \phantom{m} \\ \phantom{m} \end{matrix}$}
             & \Block{}{Markov\\ recurrence?}
             & \Block{}{Lyapunov\\ exponent?}
             & \Block{}{recurrence\\ (a.s.)?}
             & \Block{}{$\#$ visits\\ (a.s.)?}
             & \Block{}{invariant\\ density?} \\
        \Block{}{$p>1/2$}
             & \Block[fill=green!25]{}{positive recurrent}
             & \Block[fill=green!25]{}{$\begin{matrix} \phantom{m} \\ \phantom{m} \end{matrix} \lambda>0 \begin{matrix} \phantom{m} \\ \phantom{m} \end{matrix}$}
             & \Block[fill=green!25]{}{yes, in finite\\ mean time}
             & \Block[fill=green!25]{}{$ \sim n$}
             & \Block[fill=green!25]{}{supported\\ everywhere} \\
        \Block{}{$p=1/2$}
             & \Block[fill=yellow!25]{}{null-recurrent}
             & \Block[fill=yellow!25]{}{$\begin{matrix} \phantom{m} \\ \phantom{m} \end{matrix} \lambda=0 \begin{matrix} \phantom{m} \\ \phantom{m} \end{matrix}$}
             & \Block[fill=yellow!25]{}{yes, in infinite\\ mean time}
             & \Block[fill=yellow!25]{}{$ \sim n^{1/2}$}
             & \Block[fill=yellow!25]{}{infinite\\ inv.\ dens.} \\
        \Block{}{$p<1/2$}
             & \Block[fill=red!25]{}{transient}
             & \Block[fill=red!25]{}{$\begin{matrix} \phantom{m} \\ \phantom{m} \end{matrix} \lambda<0 \begin{matrix} \phantom{m} \\ \phantom{m} \end{matrix}$}
             & \Block[fill=red!25]{}{no recurrence}
             & \Block[fill=red!25]{}{$\Orb(1)$}
             & \Block[fill=red!25]{}{$\delta_0(x)$}
    \end{NiceTabular}
    \caption{Properties of the Pelikan map \eqref{eq:pelikan-closed}
    in each of its three dynamical `regimes'.}
    \label{tab:table1}
\end{table}

\subsection{Case \texorpdfstring{$\ve>0$}{ε>0}} \label{sec:F00-calculations}

We analyse our open system \eqref{eq:pelikan-markov} through the formalism of Vere-Jones \cite{vere-jones_ergodic_1967}, concerning $R$-recurrence and $R$-transience in countable absorbing Markov chains. This formalism generalises the familiar notion of positive recurrent, null-recurrent and transient Markov chains to substochastic, open and absorbing systems.
This is done through the analysis of generating functions for the iterates of the transition matrix
\begin{equation}
    G_{ji}(z) := \sum_{k=0}^\infty (M^k)_{ji}\, z^k = \sum_{k=0}^\infty \Prb(M^ki=j) \,z^k.
\end{equation}
This function has a common radius of convergence $R$ for all $i,j$, and iterates of the transition matrix obey $\lim_{n\to\infty} ((M^n)_{ji})^{1/n}=1/R$ for all $i,j$. On the radius itself, the series $G_{ji}(R) = \sum_n R^n (M^n)_{ji}$ are either all convergent or all divergent, according to which we say $M$ is $R$-transient or $R$-recurrent respectively. If $M$ is $R$-recurrent (i.e., the series diverges), we say it is $R$-null or $R$-positive as the terms of the sequence $R^n (M^n)_{ji}$ all vanish, or none of them do, as $n\to\infty$, respectively \cite{vere-jones_ergodic_1967}.

The generating function of first passage times,
\begin{equation}
    F_{ji}(z):=\sum_{k=0}^\infty \Prb\{(M^ki=j)\ \&\ (M^{k'}i\neq j \ \forall \ k'\in\{1,\dots,k-1\})\}\, z^k
\end{equation}
is related to $G_{ji}(z)$ by
\begin{equation}
    G_{ji}(z) = \begin{cases}
        1/(1-F_{ii}(z)), & i=j, \\
        G_{jj}(z)\,F_{ji}(z), & i\neq j.
    \end{cases}
\end{equation}
We can then equivalently write that $M$ is $R$-transient, resp.\ $R$-recurrent, if $F_{ii}(R) < 1$, resp.\ $F_{ii}(R) = 1$, for all $i$, and, if $F_{ii}(R)=1$, then $M$ is $R$-null, resp.\ $R$-positive, as the {derivative} $F'_{ii}(R)=\infty$, resp., $F'_{ii}(R)<\infty$ \cite{vere-jones_ergodic_1967}.

In our case,
it is quite easy to calculate
generating functions for the recurrence times of reinjection events:
consider a biased random walk on a 1D lattice which moves right with probability $p$ and left with probability $q:=1-p$, and denote the time taken to move one step to the right by the random variable $T_1$. This time must satisfy
\begin{equation}
    T_1 = \overbrace{p\cdot 1}^{\substack{\textrm{step}\\ \textrm{right}}} + \overbrace{q\cdot (1 + T_1' + T_1'')}^{\substack{\textrm{step left $\implies$ two}\\ \textrm{rightward steps needed}}}
\end{equation}
(where $T_1'$ and $T_1''$ are i.i.d.\ copies of $T_1$), and therefore the generating function of $T_1$ satisfies
\begin{equation}
    P_{T_1}(s) = p\cdot s + q\cdot s\,(P_{T_1}(s))^2 \quad \implies \quad P_{T_1}(s) = \frac{1-\sqrt{1-4pqs^2}}{2qs}
\end{equation}
for $\abs{s}\leq 1/\sqrt{4pq}$, satisfying as required\footnote{noting $\sqrt{1-4pq} = 2p-1 = p-q$} $P_{T_1}(1)=1$.
It follows that the time to take $k$ rightward steps is given by
\begin{equation}
    T_k = \sum_{i=1}^k T_1^{(i)} \quad \textrm{with} \quad P_{T_k}(s) = \left( \frac{\Delta(s)}{2qs} \right)^k, \quad \Delta:=1-\sqrt{1-4pqs^2},
\end{equation}
with each $T_1^{(i)}$ i.i.d.. A particle in state $k$ of the process \eqref{eq:pelikan-markov} needs $k$ rightward steps to reach state $0$. Since, as \eqref{eq:exp}, a successfully reinjected particle is geometrically distributed, sent to state $k$ with probability $2^{-k-1}$, then the time for the entire motion from $k$ to $0$ has generating function
\begin{equation}
    P_k(s) = \frac{1}{2-s}, \quad s<2 \quad \implies \quad
    P_T(s) = P_k(P_{T_1}(s)) = \frac{2qs}{4qs-\Delta(s)},
    \quad \abs{s}\leq\frac{1}{\sqrt{4pq}}.
\end{equation}
Finally it follows that
\begin{equation}
    F_{00}(s) = \overbrace{\frac{p(1-\ve)s}{2-P_{T_1}(s)}}^{\substack{\textrm{expansion:}\\ \textrm{reinjection, followed}\\ \textrm{by $k$ rightward steps}}}
    + \overbrace{qsP_{T_1}(s)}^{\substack{\textrm{contraction, then}\\ \textrm{one rightward step}}}
    = \frac{2pq(1-\ve)s^2}{4qs-\Delta(s)} + \frac{\Delta(s)}{2},
\end{equation}
and $G_{00}(s) = 1/(1-F_{00}(s))$.

Singularities of $G_{00}(s)$ occur in the following places: $G_{00}(s)$ has a pole when $F_{00}(s)=1$, which occurs when
\begin{equation} \label{eq:R-pos}
    \sqrt{1-4pqs^2} = ps(2-\ve)-1 \quad \iff
        \quad s=\frac{4-2\ve}{4-4p\ve+p\ve^2},
        \quad \abs{s} \leq \frac{1}{\sqrt{4pq}}.
\end{equation}
This produces a singularity in the form of a simple pole, at which $G_{00}(s)=\infty$ and $F_{00}'(s)<\infty$, at
\begin{equation} \label{eq:above}
    R_{\textrm{pole}}=\frac{4-2\ve}{4-4p\ve+p\ve^2}.
\end{equation}
If this is the unique smallest singularity,
then $R=R_{\textrm{pole}}$
and the system is $R$-positive in this region;
however, this is dependent upon the condition in \eqref{eq:R-pos}, since
there is also an algebraic branch point caused by the square root at $R_{\textrm{alg}}=1/\sqrt{4pq}$.
If
$R_{\textrm{alg}} < R_{\textrm{pole}}$, then the radius of convergence is $R=R_{\textrm{alg}}$, and the simple pole at $R_{\textrm{pole}}$ cannot occur.
In that circumstance we find
$G_{00}(R_{\textrm{alg}})<\infty$ and $F_{00}'(R_{\textrm{alg}})=\infty$, thus defining this region as $R$-transient. These two regions of parameter space meet at the boundary defined by any of the equalities\footnote{Of the four expressions in \eqref{eq:below}: this boundary occurs when $R_{\textrm{alg}} = R_{\textrm{pole}}$, producing equality between expressions 1 and 2. This occurs when the left part of \eqref{eq:R-pos} vanishes, producing the equality between 2 and 3. Expression 4 emerges from 3 if \eqref{eq:above} is written $R_{\textrm{pole}}=\frac{2(2-\ve)}{p(2-\ve)^2+4q}$.

On the boundary, the singularities of $R_{\textrm{alg}}$ and $R_{\textrm{pole}}$ collide, producing $G_{00}(R_{\textrm{alg}})=\infty$ and $F_{00}'(R_{\textrm{alg}})=\infty$, defining the system as $R$-null recurrent; however in this article we do not consider this case in any depth.}
\begin{equation} \label{eq:below}
    R=\frac{4-2\ve}{4-4p\ve+p\ve^2}=\frac{1}{\sqrt{4pq}}=\frac{1}{p(2-\ve)}=\frac{2-\ve}{4q}.
\end{equation}

In the case $\ve=0$, we find $R=1$ with $F_{00}(1)=1$ and the system is either $1$-positive or $1$-null recurrent as $p>\frac12$ or $p=\frac12$ respectively, which is consistent with findings in \cite{yan_transition_2024}.


\subsection{Principal escape rate} \label{sec:princeps}

In systems where the transfer matrix $M$ is finite,
a corollary of the Vere-Jones formalism above is that the asymptotic exponential rate of escape, and therefore the leading eigenvalue of the transfer matrix, which gives the QSD as an eigenvector, is related to $R$ by \cite{vere-jones_ergodic_1967}
\begin{equation} \label{eq:princlam}
    \lambda = 1/R \quad \textrm{and} \quad \Prb[X_n\in\Omega] \sim \lambda^n = R^{-n}.
\end{equation}
When $M$ is countably infinite, this is also demonstrated to be the case if $M$ is $R$-positive and the Markov chain is initialised from any fixed state $X_0=i$ \cite{seneta_quasi-stationary_1966}. However, as was noted in that paper, and as we will see here, the general situation is more complicated when $X_0$ is allowed to be defined by a probability distribution.
For our purposes here, we label the implied escape rate in \eqref{eq:princlam} given by $\lambda_u:=1/R$ as the `principal' escape rate.
A more thorough investigation of dynamical properties and escape rates in the system's different dynamical regimes, from selected initial conditions, was given in \cite{brevitt_weak_2026}.

\begin{figure}[ht]
    \centering
    \includegraphics[width=0.8\linewidth]{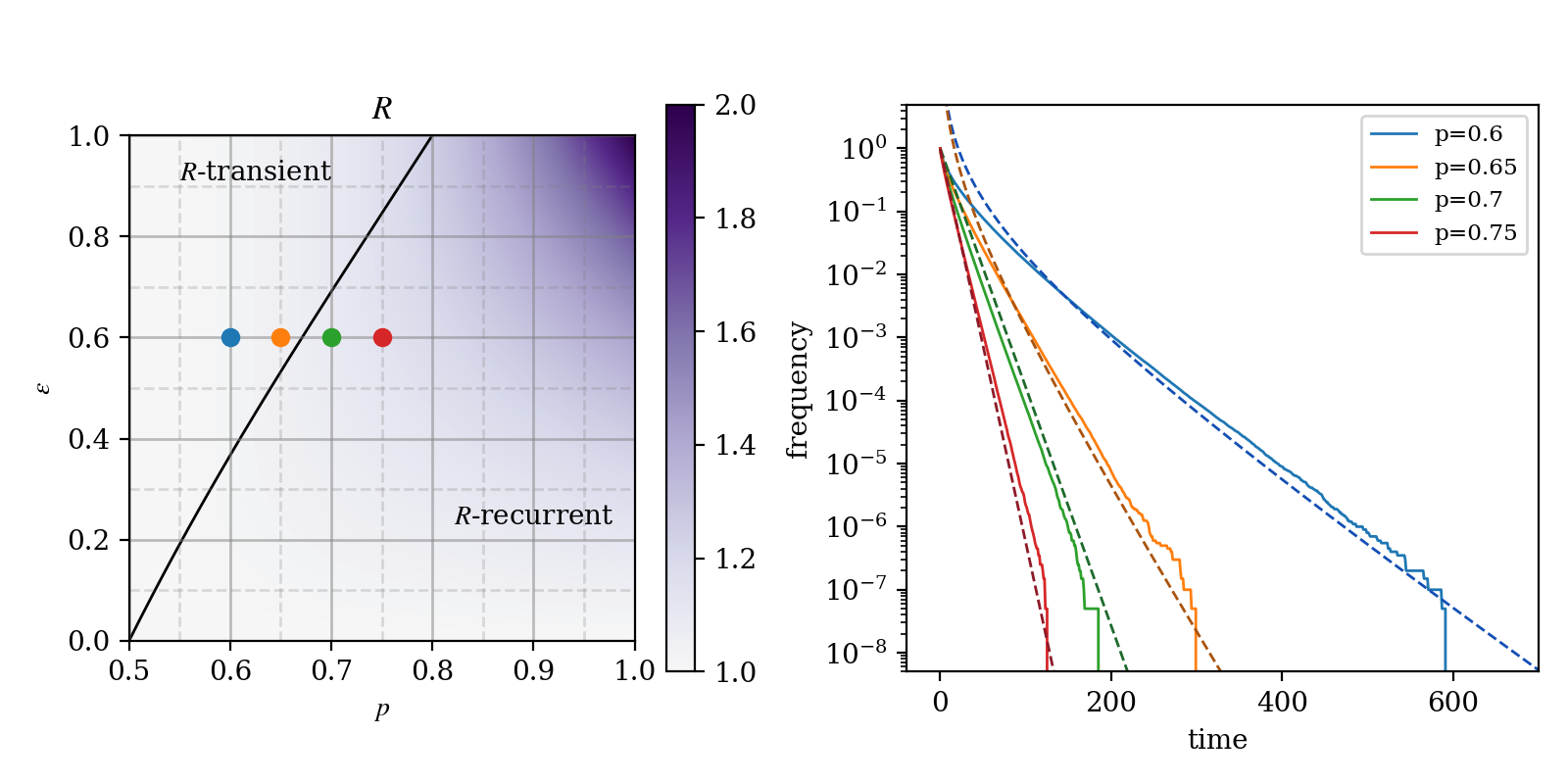}
    \caption{Left: $R$ as a function of $(p,\ve)$ shown as a colour gradient, from either side of the dividing line \eqref{eq:below}.
    Right: Survival probability $\Prb[X_n\in\Omega]$ as a function of time $n$, obtained from simulations (solid lines), plotted against predictions obtained from \eqref{eq:princlam} (dashed lines), for the four identified sample points shown left. Sample size $N=2\times10^7$. In the $R$-transient regime, simulations are affected by a non-exponential transient, see \cite{brevitt_weak_2026} for details.}
    \label{fig:powerlaw}
\end{figure}

Among other things, we note that $1\leq R\leq 2$, increasing monotonically with $p$ and $\ve$, achieving its maximum at $(p,\ve)=(1,1)$, implying a maximum escape rate of $\Prb[X_n\in\Omega]\sim 2^{-n}$. We plot $R$ as a colour gradient in Fig.~\ref{fig:powerlaw}, showing the two dynamical regimes of $R$-recurrence ($R$-positivity) and $R$-transience, as a function of $p$ and $\ve$. We also test escape rates against simulations, where they are found to match well.

\section{Conditionally invariant measures (QSDs)} \label{sec:2}

\subsection{Analytic calculations} \label{sec:2.1}

In this section we analytically calculate the QSDs of our system.
Let $\mu_k(t)$ denote the mass of a probability measure $\mu(t)$ (not necessarily invariant) in the state $k$ of the Markov chain \eqref{eq:pelikan-markov} at some time $t$, normalised so that $\abs{\mu(t)}=\sum_{k=0}^\infty \mu_k(t)=1$. Under the evolution of the system,
\begin{equation} \label{eq:proevo}
    \mu_k(t+1) = M[\mu(t)]_k = \begin{cases}
        {p\,\mu_{k+1}(t)}
        + {\mu_0(t)\, 2^{-k-1}\, p(1-\ve)},
        & k>0, \\
        {p\,\mu_1(t)} + {\frac12 \mu_0(t)\, p (1-\ve)}, & k=0.
    \end{cases}
\end{equation}
Let $A_t(z) := \sum_{k=0}^{\infty} \mu_k(t)\, z^k$. Then $A_t(z)$ evolves in time under the master equation
\begin{equation} \label{eq:masteq}
    A_{t+1}(z)=M[A_t](z) = {\frac{A_t(z)-\mu_0}{z}p} + {zqA_t(z)} + {\frac{\mu_0(t)\,p(1-\ve)}{2-z}},
\end{equation}
Note that $A_t(1) = \sum_{k=0}^\infty \mu_k(t)$ defines the normalisation of the measure $\abs{\mu(t)}$, and this master equation satisfies
\begin{align} \label{eq:normalisation}
    M[A_t](1)
        &= A_t(1)\,[p+q] - p\,\mu_0(t) + \mu_0(t)\,p(1-\ve) \nonumber \\
        &= 1-p\ve\, \mu_0(t),
\end{align}
meaning that at each timestep, a proportion $p\ve\mu_0(t)$ of the system escapes. This is entirely consistent with our {\it a priori} dynamical understanding of the system.\footnote{Consider that a particle has $X_t=0$ with probability $\mu_0(t)$, applies $M_\mathrm{exp}$ with probability $p$, and under those circumstances is killed with probability $\ve$.}
Then any QSDs of the system will be given by eigenfunctions $A(z)$ satisfying $M[A]=\lambda A$, and \eqref{eq:normalisation} implies $\lambda = 1-p\ve\mu_0$, i.e.:
\begin{gather}
    (1-p\ve \mu_0)A = M[A] = \left(\frac{p}{z}+zq\right)A - \frac{p\mu_0}{z} + \frac{p\mu_0(1-\ve)}{2-z} \nonumber \\
    \implies A(z) = p \mu_0 \frac{2-(2-\ve)z}{(2-z)(z^2q + (p\ve \mu_0 - 1)z + p)}.  \label{eq:Az}
\end{gather}
We note the following constraints on $A(z)$:
\begin{itemize}
    \item $\mu_0$ is a free variable, since it appears in the solution for which we are trying to solve, and $\lambda$ is a function of $\mu_0$;
    \item $A(0)=\mu_0$, for internal consistency;
    \item $A(1)=\sum_{k=0}^{\infty} \mu_k = 1$;
    \item $\mu_k \geq 0 \ \forall \ k\in\N\cup\{0\}$. 
\end{itemize}
We have already addressed the normalisation, and indeed one can verify
\begin{equation}
    A(1) = p\mu_0 \frac{2-(2-\ve)}{q+p\ve\mu_0-1+p}
     = p\mu_0 \frac{\ve}{p\ve\mu_0} = 1
\end{equation}
whenever $\ve>0$ (the case $\ve=0$ is discussed below); and we also find
\begin{equation}
    A(0) = p \mu_0 \frac{2}{2p} = \mu_0
\end{equation}
as we should. In the case $\ve>0$, the final constraint is more complicated, and will be dealt with separately below.

We now consider separately the cases $\ve=0$ and $\ve>0$.

\subsubsection{Case \texorpdfstring{$\ve=0$}{ε=0}.} \label{sec:recreating-jin}

When $\ve=0$,
corresponding to the closed system \eqref{eq:pelikan-closed-markov}, for which results are already known \cite{yan_transition_2024}, $A(z)$ reduces to
\begin{equation}
    A(z) = p \mu_0 \frac{2-2z}{(2-z)(z^2q-z+p)} = \frac{2p\mu_0}{(2-z)(p-qz)}.
\end{equation}
In this case, the invariant measure is determined completely by imposing the normalisation
\begin{equation}
    A(1) = \frac{2p\mu_0}{p-q} = 1 \implies \mu_0 = \frac{2p-1}{2p}
\end{equation}
which agrees with \eqref{eq:invmeas} and \cite{yan_transition_2024}. Substituting this in, we obtain
\begin{equation}
    A(z) = \frac{2p-1}{(2-z)(p-qz)}
    = \frac{2p-1}{3p-2} \left[ \frac{1}{2-z} - \frac{1}{\frac{p}{q}-z} \right]
\end{equation}
implying
\begin{equation}
    \mu_k = \frac{2p-1}{3p-2} \left[ 2^{-k-1} - \left(\frac{q}{p}\right)^{k+1} \right]
\end{equation}
exactly matching \eqref{eq:invmeas}, as it should.


\subsubsection{Case \texorpdfstring{$\ve>0$}{ε>0}.}

In the general case $\ve>0$, however, we are not so lucky; we are unable to further constrain our solutions $A(z)$, and the quantity $\mu_0$ remains in the equations as a free variable. For the avoidance of confusion, let us rename this free parameter $\alpha$. Then both ${\mu} = (\mu_0, \mu_1, \dots)$ and $\lambda$ are dependent on $\alpha$, with in particular
\begin{equation} \label{eq:princalp}
    \lambda(\alpha)=1-p\ve\alpha.
\end{equation}
Given a value of $\alpha$, individual terms $\mu_k$ can be computed from $A(z)$ via a recurrence relation: since \emph{in general}
\begin{align}
    A(z) &= P(z) / [1+q_1z+q_2z^2+q_3z^3] \nonumber \\
    \implies A(z) &= -q_1zA(z) -q_2z^2A(z) -q_3z^3A(z) + P(z) \nonumber \\
    \implies \mu_k &= -q_1 \mu_{k-1} -q_2 \mu_{k-2} -q_3 \mu_{k-3} + p_k, \qqqqquad k\geq 3,
\end{align}
we therefore write \emph{in our case}
\begin{equation}
    A(z) = \frac{\alpha(1-z\,\frac{2-\ve}{2})}{1 - \frac{2\lambda+p}{2p}z + \frac{2q+\lambda}{2p}z^2 - \frac{q}{2p}z^3}
\end{equation}
implying
\begin{equation}
    \mu_k = \frac{1}{2p} \left[ (2\lambda+p) \mu_{k-1} - (2q+\lambda) \mu_{k-2} + q \mu_{k-3} \right] + p_k
\end{equation}
where $p_0 = \alpha$, $p_1 = \frac12 \alpha(2-\ve)$, and $p_k=0$ for $k\geq 2$. The first few terms can be computed explicitly, using $\mu_k = \frac{1}{k!}A^{(k)}(0)$, as
\begin{align} \label{eq:recrel1}
    \mu_0 &= A(0) = \alpha \nonumber \\
    \mu_1 &= A'(0) = \frac{\alpha}{2p}(2\lambda -p + p\ve) \nonumber \\
    \mu_2 &= \frac12 A''(0) = \frac{\alpha}{4p^2} [4(\lambda^2-pq) - p(1-\ve)(p +2\lambda) ]
\end{align}
and all subsequent terms by the recurrence relation (from \eqref{eq:recrel1}, since $p_k=0$)
\begin{equation} \label{eq:recrel}
    \mu_k = \frac{1}{2p} \left[ (2\lambda+p) \mu_{k-1} - (2q+\lambda) \mu_{k-2} + q \mu_{k-3} \right].
\end{equation}
This is the method we have used to numerically calculate and plot the QSDs in all following figures.

\subsection{Analysis of positivity} \label{sec:bddns-qsd}

While we have easily been able to apply the constraints $A(0)=\mu_0$ and $A(1)=1$ to our solutions, it has not removed our free parameter in the case $\ve>0$. We do also need to consider the third constraint, $\mu_k\geq0$ for all $k\geq0$. This, it turns out, is very difficult to determine by inspection of the generating function alone.

The asymptotic rate of decay of $\mu_k$ as $k\to\infty$ (the `shape' of the distribution) can be determined by the singularities of $A(z)$ \cite{flajolet_analytic_2009}, which here are the zeroes of the denominator $(2-z)(z^2q - \lambda z + p)$, with each root $r_i$ contributing an exponential term
proportional to $r_i^{-k}$; namely,
\begin{equation}
    A(z) = \sum_i \frac{B_i}{z - r_i} \quad \implies \quad \mu_k = \sum_i C_i\,r_i^{-k}, \quad C_i=-B_i/r_i.
\end{equation}
The qualitative form of the solutions therefore depends strongly on the polarity of the roots $r_i$: if all roots are real, positive, and greater than one, the solution will be a mixture of exponential decays. On the other hand, if any $r_i$ is complex, negative, or less than one, it will necessarily contribute complex, oscillatory, or exponentially growing terms, respectively.
Roots of the denominator are given by $z=2$ and by the roots of $(z^2q - \lambda z + p)$, the discriminant of which is $\Delta = \lambda^2-4pq$, which has its critical point at
\begin{equation} \label{eq:a0s}
    \lambda^2 -4pq = 0 \implies (1-p\ve\alpha)^2 = 4pq \implies \alpha_c = \frac{1-\sqrt{4pq}}{p\ve}.
\end{equation}
If $\alpha>\alpha_c$, then $(z^2q-\lambda z+p)$ has complex roots, which produce oscillatory densities, which we can immediately exclude. Otherwise, roots are given by
\begin{equation}
    z = \frac{\lambda \pm \sqrt{\lambda^2 - 4pq}}{2q} \quad \textrm{and} \quad z=2.
\end{equation}
Further, even if the $r_i$ are all real and $r_i>1$, it still frequently occurs that some of the coefficients $C_i$ of these exponential decays are negative, in which case $\mu_k$ may, or may not, dip below zero.
It is therefore necessary for us to determine precisely the range of $\alpha$ for which all ${\mu_k}\geq0$, as only these correspond to valid QSDs.

Fortunately, we are saved by a degree of regularity:
in all cases, the asymptotic (large $k$) shape is determined by the slowest decaying exponents, i.e., those with minimal $\abs{r_i}$. This offers us some guarantees in the large $k$ limit: name this minimal root $r_1$, and assume it is real, greater than one, unique, and has coefficient $C_1>0$;
then $r_1^{-k} \gg \abs{r_{i>1}^{-k}}$ asymptotically, implying
\begin{equation} \label{eq:condition}
    C_1 r_1^{-k} > \sum_{i>1} \abs{C_i r_i^{-k}} \quad \textrm{for all $k\geq k^*$,} \quad \textrm{for some $k^*$.}
\end{equation}
Hence, beyond this point $k^*$, $\mu_k > 0$ if and only if $C_1 r_1^{-k} > 0$. Therefore, to demonstrate $\mu_k\geq 0$ everywhere, it is only really necessary to demonstrate that
$\mu_k\geq 0$ for all $k$ \emph{up until this criterion is reached}, since beyond that point we are assured of it. This therefore enables us to \emph{rigorously} determine the non-negativity of $\mu$ numerically in finite time.

\begin{figure}[ht]
    \centering
    \includegraphics[width=0.98\linewidth]{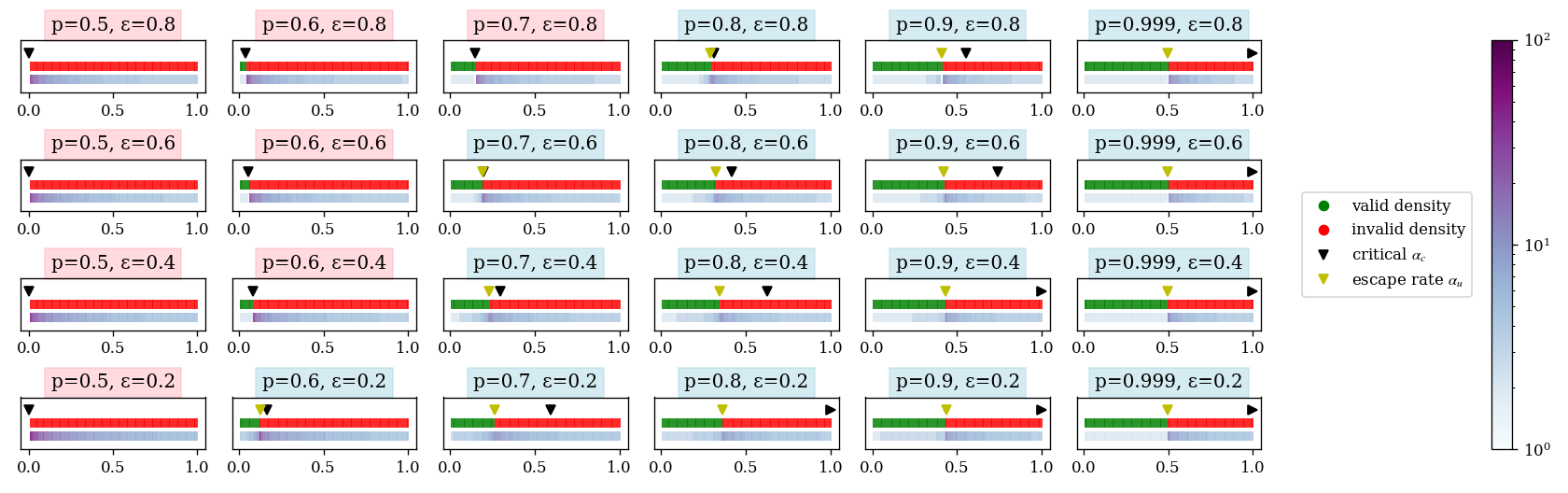}
    \caption{`Valid' and `invalid' values of $\alpha$, for a range of $p,\ve$. Values of $\alpha$ producing $\mu_k \geq 0$ for all $k$ are shown in green. Invalid $\alpha$, for which $\mu_k < 0$ for some $k$, are shown in red.
    Plot titles are coloured to indicate the 
    dynamical regions discussed in Sec.~\ref{sec:F00-calculations}, with $R$-recurrent parameters shown in blue and $R$-transient parameters in red.
    The critical value $\alpha_c$ is shown in black, and, where it differs from this, the asymptotic $\alpha_u$ derived from the principal escape rate is shown in gold.
    Below each subplot: a bar indicating the number of iterations required to determine validity/invalidity, according to the colour scale shown on the right of the figure.}
    \label{fig:fig1-vinvdens}
\end{figure}

In Fig.~\ref{fig:fig1-vinvdens} we show the range of valid and invalid
$\alpha\in [0,1]$, for each $(p,\ve)$, colour coded by whether or not the corresponding ${\mu}\geq 0$.
This was determined by numerically calculating $\mu_k$
from the recurrence relation \eqref{eq:recrel1} and \eqref{eq:recrel}, and testing for non-negativity,
increasing $k$ until one of two conditions was met: either
\begin{itemize}
    \item $\mu_k<0$, in which case the distribution is marked as invalid, and the corresponding $\alpha$ is shown in red in Fig.~\ref{fig:fig1-vinvdens}; or
    \item $k>k^*$ according to the criterion \eqref{eq:condition} (the coefficients $C_i$ were determined numerically from $A(z)$ in each case), in which case the distribution is marked as valid, and the corresponding $\alpha$ is shown in green in Fig.~\ref{fig:fig1-vinvdens}.
\end{itemize}
Further, below each bar, we plot by means of a colour gradient the number of iterations $k$ required for this to occur, which one can see is generically small.
Above each bar, the location of the critical $\alpha_c$ is shown in black, if $\alpha_c<1$. We also plot in gold the value $\alpha_u$ implied by the principal escape rate:
from \eqref{eq:princlam}, we identify the principal $\lambda_u := 1/R$, and therefore define the principal $\alpha_u$ such that $\lambda_u=1-p\ve\alpha_u$, per \eqref{eq:princalp}.
This eigenvalue is $\lambda_u = (4-4p\ve+p\ve^2)/(4-2\ve)$
in the case of $R$-positivity,
implying $\alpha_u = (4p-2-p\ve)/(2p(2-\ve))$,
and $\lambda_u=\sqrt{4pq}$
in the case of $R$-transience,
implying $\alpha_u = (1-\sqrt{4pq})/p\ve$.
Generically, $\alpha_u<\alpha_c$ in the region of $R$-positivity, and $\alpha_u=\alpha_c$ in the region of $R$-transience.

Fig.~\ref{fig:fig1-vinvdens} confirms that, when $\alpha>\alpha_c$, the produced distributions are always invalid; this is because of the oscillatory problem discussed above. For $\alpha<\alpha_c$, we may or may not have a valid distribution; we rather observe numerically that they are valid for $\alpha\leq\alpha_u$ and invalid otherwise. A corollary of this is that $\lambda_u$ is the minimal possible eigenvalue (meaning, the fastest possible escape rate), and all $\lambda>\lambda_u$ are supported by some $\alpha$.

\subsection{Verification in simulations} \label{sec:qsd-simulations}

\begin{figure}[ht]
    \centering
    \includegraphics[width=0.8\linewidth]{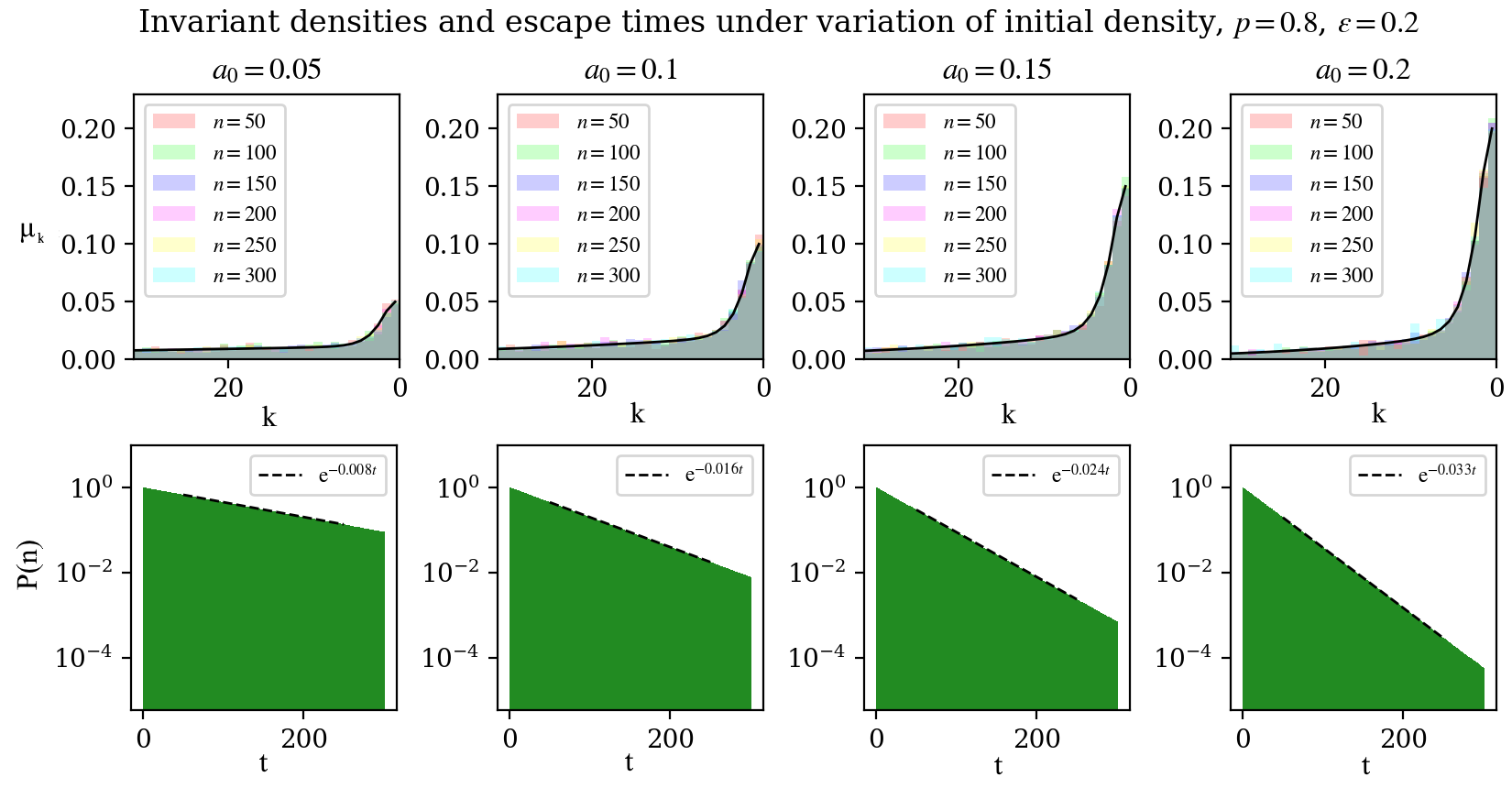}
    \caption{Top: Histograms of QSDs for a variety of $\alpha$, for the sample parameters $(p,\ve)=(0.8,0.2)$, which are preserved under the time evolution \eqref{eq:proevo}, normalised. The index $k$ of each state is plotted from right to left, with the height of the bar indicating the probability mass $\mu_k$. QSDs are evolved under \eqref{eq:proevo} for $n$ timesteps, for a variety of $n$, in varying colours, which are overlaid, thus producing grey bars where they overlap. Our theoretical QSD for each chosen value of $\alpha$, produced by \eqref{eq:recrel1} and \eqref{eq:recrel}, is shown in a solid black line in each case. Bottom: In each case, the surviving population $\Prb[X_n\in\Omega]$ as a function of time $n$; the exponential escape rate \eqref{eq:princlam} from our predictions is plotted in dashed black.}
    \label{fig:osci2}
\end{figure}

We can verify in simulations that this plurality of QSDs is genuine by, for each $\alpha$, initialising our system from its corresponding QSD, and evolving the system with time to observe that the initial measure is preserved, which we see in Fig.~\ref{fig:osci2}. Further, each of the measures is shown to produce a different exponential escape rate, corresponding to the different $\lambda(\alpha)$, exactly as predicted.

\section{Analysis of shape} \label{sec:shape}

\subsection{Existence of QSDs of specified shapes}

We make a brief analysis of the asymptotic `shape' of the QSDs $\mu_k$ with respect to $k$. This turns out to be important, because we will show in this section that the asymptotic shape is a conserved quantity of the system dynamics. We focus especially on the QSDs given by the `principal' choices $\lambda_u$ and $\alpha_u$, for which we give explicit formulae.

For some parameter values $(p,\ve)$, the system supports a `principal' QSD (given by $\alpha=\alpha_u$) which has $\mu_k\sim 2^{-k}$, while for other parameters the rate of decay of $\mu_k$ is slower. This is significant because, when considered on the partition \eqref{eq:partition}, a distribution with $\mu_k\sim2^{-k}$ is bounded and has bounded variation on $[0,1]$, and not otherwise. This is an important feature in dynamical systems theory, see \cite{brevitt_weak_2026, bose_ulams_2013} for further discussion. This occurs when the pole at $z=2$ is the smallest singularity of $A(z)$ \eqref{eq:Az}. In both cases, non-principal QSDs with lesser values of $\alpha$ necessarily correspond to greater $\lambda$, slower rates of escape, and more slowly decaying shapes $\mu_k$.

Inserting $\lambda_u$ into $A(z)$ obtains the generating function for the principal QSD $\mu^*$ in the $R$-positive regime as
\begin{equation} \label{eq:pfdecay-1}
    A(z) = p\alpha_u \frac{2-(2-\ve)z}{(2-z)(z^2q-\lambda_u z+p)} = \frac{2-4p+p\ve}{2q(2-z)(z-\frac{p(2-\ve)}{2q})}
\end{equation}
and in the $R$-transient regime as
\begin{equation} \label{eq:pfdecay-2}
    A(z)
    = \frac{q(1-\sqrt{4pq})(2-(2-\ve)z)}{\ve(2-z)(z- \sqrt{p/q} )^2}.
\end{equation}
We see that $z=2$ is the smallest pole if $4q < p(2-\ve)$ in the $R$-positive regime; otherwise, and in the entirety of the $R$-transient regime, there is no value of $\alpha$ producing a QSD of this shape. The line separating these two cases is shown in Fig.~\ref{fig:map} and Fig.~\ref{fig:color-plot-6-parts} (left) below, where in the latter the asymptotic `shape' is also shown as a colour gradient. This line intersects $\ve=0$ at $p=\frac23$, which exactly matches to results found in \cite{yan_transition_2024}. The shape of the principal QSD will become significant when we consider the existence of QEDs in Sec.~\ref{sec:qed}.

\subsection{Shape as a conserved quantity} \label{sec:quant}

In this section we demonstrate the conservation of the asymptotic `shape' of a distribution $\mu_k$ under the system dynamics. This will be an important factor to consider when assessing the stability of various QSDs.

This section will use the notation of Flajolet and Sedgewick, that we write $a_n \bowtie K^n$ if
$\limsup_{n\to\infty} \abs{a_n}^{1/n} = K$ \cite{flajolet_analytic_2009}.
This relation is {strictly weaker} than $a_n \sim K^n$.
This notation allows the important result for generating functions that if $A(z) = \sum_{k=0}^{\infty} a_k z^k$ has radius of convergence $\varrho(A)$, then
\begin{equation}
    \varrho(A) = 1/K \quad \iff \quad a_k \bowtie K^n.
\end{equation}

Consider the normalised transfer operator of the dynamics on a generic (not necessarily invariant) distribution $\mu$ (cf.\ \eqref{eq:proevo}),
\begin{equation} \label{eq:transfer}
    (M^*[\mu])_k = \frac{1}{1-p\ve\mu_0} \begin{cases}
        {\mu_{k+1} p} + {\mu_{k-1} q} + {\mu_0 p 2^{-k-1} (1-\ve)}, & k>0, \\
        {\mu_1 p} + {\frac12 \mu_0 p (1-\ve)}, & k=0.
    \end{cases}
\end{equation}
It is straightforward to see that, if the `shape' of $\mu_k$
decays exponentially or slower with respect to $k$ (so that $\mu_{k-1} \bowtie \mu_k \bowtie \mu_{k+1}$), then
\begin{equation} \label{eq:Msim}
    (M^*[\mu])_k \bowtie {\mu_{k+1}} + {\mu_{k-1}} + {2^{-k}} 
    \bowtie \mu_k + 2^{-k}, \quad k\to\infty.
\end{equation}
From this we draw the following conclusions:
\begin{itemize}
    \item If $\mu_k \bowtie \e^{-\sigma k}$, $0\leq \sigma < \ln 2$,
    then $(M^*[\mu])_k$ will also have $(M^*[\mu])_k \bowtie \e^{-\sigma k}$;
    \item If $\mu_k \bowtie \e^{-\sigma k}$, $\sigma \geq \ln 2$, or $\mu_k \bowtie 0$, then $(M^*[\mu])_k$ will have $(M^*[\mu])_k \bowtie 2^{-k}$.
\end{itemize}
This argument can also be formalised in terms of the generating function $A(z):= \sum_{k=0}^{\infty} \mu_k z^k$, where $\sigma$ as above corresponds directly to the radius of convergence $\varrho$ of $A(z)$,
by $\varrho(A) = e^\sigma$ (i.e., $\mu_k \bowtie \e^{-\sigma k} = \varrho^{-k}$). We then see that the normalised transfer operator on the generating function (cf.\ \eqref{eq:masteq})
\begin{equation}
    M^*[A](z) = \frac{1}{1-p\ve\mu_0} \left(
        {\frac{A-\mu_0}{z}p} + {zqA} + {\frac{p\mu_0(1-\ve)}{2-z}}
    \right)
\end{equation}
introduces a pole at $z=2$, and preserves existing poles elsewhere, implying the same results as above -- the closest pole to the origin, which dictates the asymptotic shape of $\mu_k$, will either be preserved if less than $2$, or $2$ otherwise.

Hence, we conclude that the `shape' of $\mu_k$ as $k\to\infty$ is, within these equivalence classes, a \emph{conserved quantity} of the dynamics,
which divides
distribution space into uncountably many \emph{invariant sets} indexed by $\sigma$.
Since, in our QSDs, the shape $\sigma$ is an injective function of $\alpha$, for any given $(p,\ve)$, each equivalence class contains \emph{at most} one QSD.
We therefore conjecture that, if the process is initialised with an initial distribution $\mu$ in equivalence class $\sigma$, and there exists a QSD $\mu^*$ within the same equivalence class $\sigma$, then $(M^*)^n[\mu] \to \mu^*$.
On the other hand, if there is no existing QSD within the class, we have a more complex situation. We investigate both scenarios further in Sec.~\ref{sec:stability}.

\section{Stability analysis of QSDs} \label{sec:stability}

The purpose of this section is to analyse the stability of all our QSDs, both principal and non-principal,
to perturbation by random noise.
Stable QSDs would represent steady states which have the ability to greatly influence the dynamics of generic trajectories; whereas unstable QSDs can be discarded as significantly less dynamically important.
Our results in Sec.~\ref{sec:quant} provide us with broad predictions for the limiting QSD of a chosen initial measure $\mu$, according to the measure's asymptotic shape.
Our scrutiny will be primarily focused on the non-principal QSDs, as these are the ones which are not as well accounted for in the literature \cite{seneta_quasi-stationary_1966}, not as easily detected by numerical methods such as the Ulam method \cite{brevitt_weak_2026, bose_ulams_2013}, and for which other results such as corresponding QEDs are unavailable (Sec.~\ref{sec:qed}).

\subsection{Numerical testing} \label{sec:testing}

Since we are operating in the infinite dimensional Banach space $\mu\in\ell_1$, usual stability methods concerning the spectrum of the transfer operator no longer work, especially since the operator is not compact and its spectrum is not discrete. Instead, therefore, we consider the stability of QSDs against perturbation by random noise in $\ell_1$.
This will be done using variations on the following procedure:


\begin{figure}[ht]
    \centering
    \begin{minipage}[t]{0.48\textwidth}
        \centering
        \includegraphics[align=c,width=0.8\linewidth]{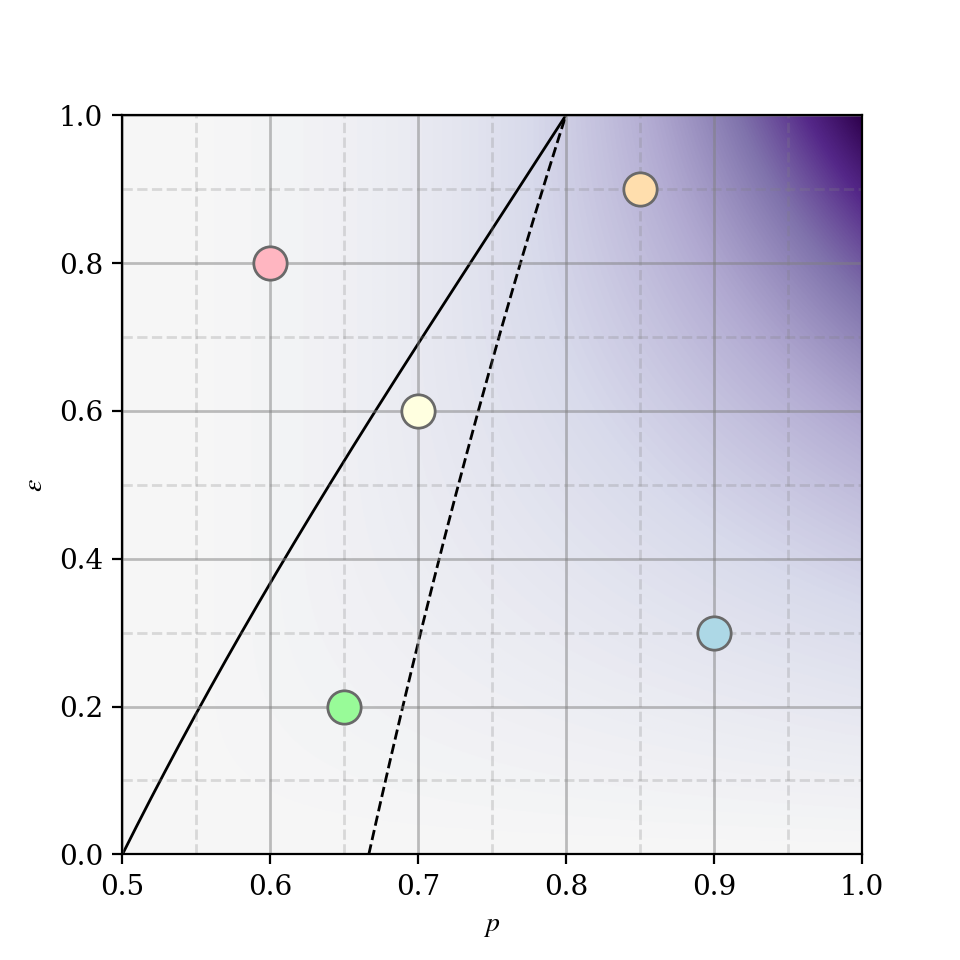}
    \end{minipage}
    \hfill
    \begin{minipage}[t]{0.48\textwidth}
        \centering
        \begin{NiceTabular}[hvlines]{ccc}
            & \Block{}{$R$-positive?} & \Block{}{$\mu_k\sim 2^{-k}$?} \\
            \Block[fill=r]{}{red point} & \n & \n \\
            \Block[fill=g]{}{green point} & \y & \n \\
            \Block[fill=w]{}{white point} & \y & \n \\
            \Block[fill=o]{}{orange point} & \y & \y \\
            \Block[fill=b]{}{blue point} & \y & \y
        \end{NiceTabular}
    \end{minipage}
    \caption{Left: A map showing the five sample points in parameter space used in Sec.~\ref{sec:testing}.
    Regions of $R$-recurrence and $R$-transience are demarcated by a solid line, while regions where the principal QSD has $\mu_k\sim2^{-k}$ or not are demarcated by a dashed line. The colour gradient shows the value of $R$,
    which is also the principal escape rate. Right: Properties of the five identified sample points.
    }
    \label{fig:map}
\end{figure}

\begin{figure}[ht]
    \centering
    \includegraphics[width=0.8\linewidth]{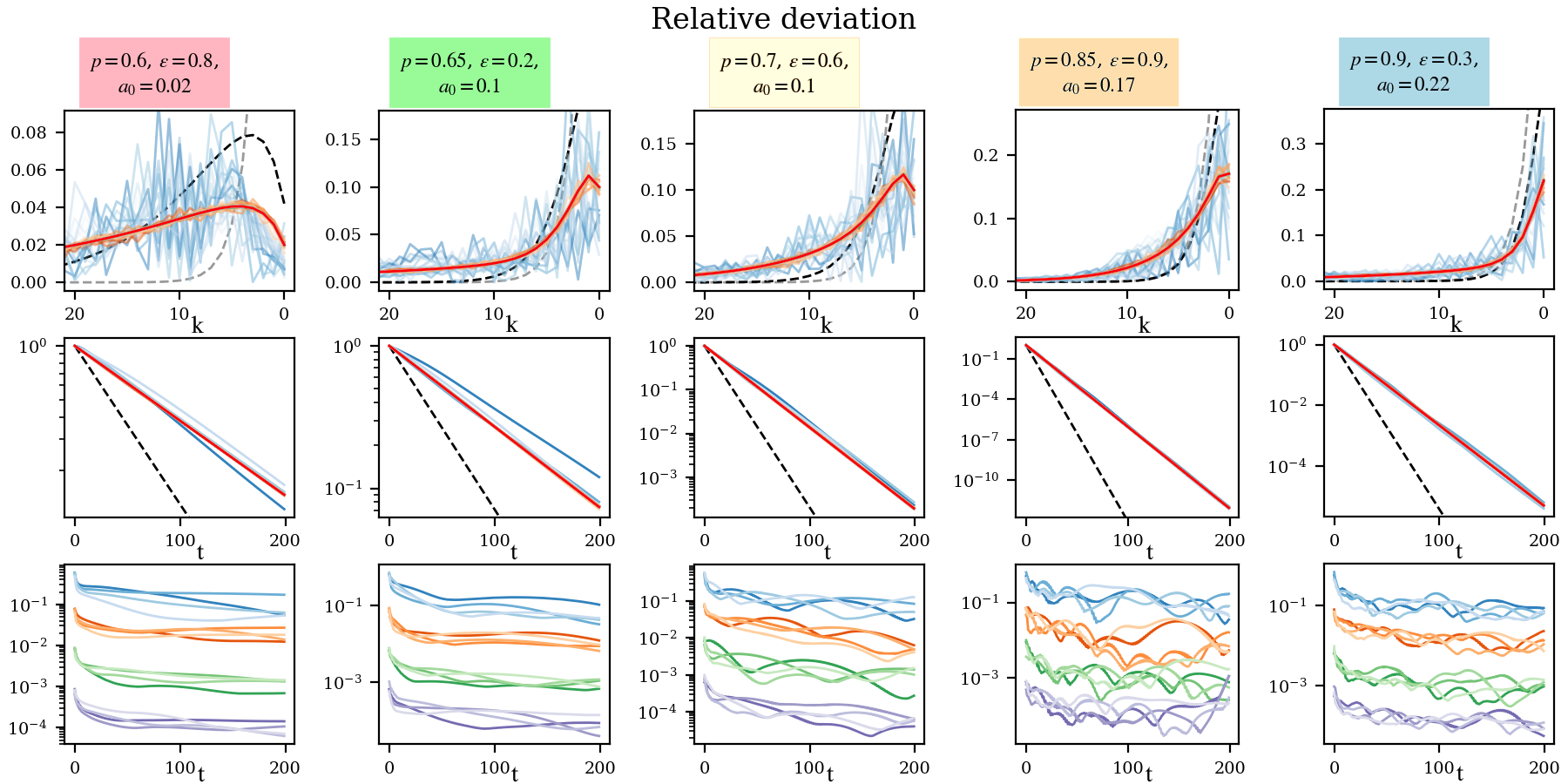}
    \caption[Stability analysis of non-standard QSDs in the open Pelikan map]{Time evolution of perturbations of non-standard QSDs, for a variety of $p,\ve$, see text for details.}
    \label{fig:test2}
\end{figure}

\begin{itemize}
    \item We choose five sample points $(p,\ve)$ in parameter space to sample. These points are chosen to
    cover as wide an area of parameter space as possible, and are shown in Fig.~\ref{fig:map}.
    \item For each chosen $(p,\ve)$, we identify both the \emph{principal} QSD $\mu^*$ (shown in dashed black in Fig.~\ref{fig:test2}), and a chosen \emph{non-principal} QSD $\mu^\dagger$ (shown in solid red), parametrised by a different value of $\alpha$, which will be given in the subplot title.
    \item We produce 16 random perturbations $\mu^{\dagger\#}$ of the non-principal QSD $\mu^\dagger$: four i.i.d.\ random perturbations each for four different amplitudes of random noise (shown in varying shades of -- in order of decreasing magnitude of noise -- blue, orange, green, and purple). Note that in Fig.~\ref{fig:test2}, not all perturbations are always visible, due to their small magnitudes of noise.
    \item We then iteratively evolve each perturbed distribution using the normalised transfer operator equation \eqref{eq:transfer}. We can track the behaviour of each evolved distribution $(M^*)^n[\mu^{\dagger\#}]$ in comparison with $(M^*)^n[\mu^{\dagger}]=\mu^\dagger$ and $(M^*)^n[\mu^*]=\mu^*$ via the following methods (shown in Fig.~\ref{fig:test2} from top to bottom):
    \begin{itemize}[label=\textbullet]
        \item visual inspection of the evolved distribution $(M^*)^n[\mu^{\dagger\#}]$, by comparison with the unperturbed distribution $\mu^\dagger$ and the principal QSD $\mu^*$, in the top row of plots -- in Fig.~\ref{fig:test2}, only the largest perturbations are however clearly visible;
        \item deviation of the escape rate: the central row of plots show the surviving population of the system
        on a semi-log axis as a function of time, the gradient of which indicates the asymptotic escape rate. This is shown for the non-principal QSD $\mu^\dagger$ in red, the principal QSD $\mu^*$ in dashed black, and the perturbations $\mu^{\dagger\#}$ in their allocated colours. Since there is a connection between the choice of QSD $\mu$ and the escape rate $\lambda$, the limiting asymptotic escape rate is an indicator of the limiting QSD;
        \item the $\ell^1$ distance between
        $(M^*)^n\mu^{\dagger\#}$ and the unperturbed non-principal QSD $\mu^\dagger$,
        shown in the bottom row on a semi-log axis, as a function of time: if $\mu^\dagger$ is stable, this should decrease to zero, while on the other hand the maximum possible $\ell_1$ distance between two normalised distributions is $2$. The initial $\ell^1$ distance between $\mu^{\dagger\#}$ and $\mu^\dagger$ is given by the size of the perturbation.
    \end{itemize}
\end{itemize}
For the purposes of the computer, distributions are stored as $1000$-dimensional vectors $[\mu_0,\dots,\mu_{999}]$, with higher $\mu_k$ truncated. We typically evolve for $200$ timesteps, so boundary effects from the truncation will not affect our results.

There are many ways to generate random perturbations of an initial distribution (and we tried several for comparison), but here we show one: we perturb each element $\mu^\dagger_k$ relative to its initial size, by
\begin{equation}
    \mu_k^{\dagger\#} \propto \mu^\dagger_k \cdot \abs{1 + \epsilonhat \xi_k} \quad \textrm{for} \quad \xi_k \sim \mathcal{N}(0,1) \ \textrm{i.i.d.}.
\end{equation}
This is then normalised to produce the perturbed distribution $\mu^{\dagger\#}$.
In Fig.~\ref{fig:test2} this is performed for $\epsilonhat \in \{ 10^{0}, 10^{-1}, 10^{-2}, 10^{-3} \}$. We observe that the escape rate remains consistent with that of the unperturbed, non-standard distribution $\mu^\dagger$ (red line), while the $\ell_1$ distance from $\mu^\dagger$ remains more or less constant, decreasing if anything only slightly. Rather, what we see is a sort of quasi-stable cyclic fluctuation or oscillation around the initial perturbed distribution $\mu^{\dagger\#}$, seemingly without clearly converging or diverging.
Nevertheless we observe that the shape $\mu_k^{\dagger\#} \bowtie \e^{-\sigma k}$ is conserved, and the escape rate is also seen to be extremely robust.

Examples of different modes of convergence seen for other types of noise that we tested (for which we neglect to show figures) include:
\begin{itemize}
    \item `shaped' additive noise:
    naïvely, we may wish to apply additive noise of constant magnitude, $\epsilonhat\xi_k$ with $\xi_k\sim\Norm(0,1)$ i.i.d. on each $\mu^\dagger_k$ and $\epsilonhat$ constant.
    This is generally not possible due to problems of non-negativity and normalisation in the tail as $k\to\infty$.
    Instead, we must `shape' our noise in order to guarantee convergence, which we do in two ways below:
    \begin{itemize}[label=\textbullet]
        \item `thick' noise, tapered by some slowly decaying function, e.g.\
        $\mu_k^{\dagger\#} \propto \abs{\mu^\dagger_k + \epsilonhat_k \xi_k}$, for $\epsilonhat_k \propto k^{-2}$, $\xi_k \sim \Norm(0,1)$ i.i.d.;
        then one obtains $\mu_k^{\dagger\#} \bowtie 1$ a.s.,
        and in simulations we see convergence to a wide, flat, almost uniform distribution, reminiscent of an infinite invariant density on the half-line, which does not correspond to any QSD, and which produces in simulations a power-law escape rate. The dynamics of an infinite ergodic scenario on the open Pelikan map would be without doubt interesting and important,
        but in this paper we do not consider this case.
        \item `slim' noise, tapered by an exponential function decaying at least as fast as the QSD, e.g.\
        $\epsilonhat_k \propto 2^{-k}$;
        this does not affect the overall `shape' of the distribution and therefore $\mu_k^{\dagger\#} \bowtie \mu_k^\dagger$.
        In simulations we find $(M^*)^n[\mu^{\dagger\#}]\to \mu^\dagger$ at an exponential rate, in the $\ell_1$ norm, with respect to time $n$.
    \end{itemize}
    \item a sampled distribution, which we believe may be relevant for several physical and computational scenarios: we take $\mu_k^{\dagger\#}$ to be the empirical distribution of an ensemble of size $N$ sampled from the distribution $\mu_k^\dagger$; thus, each $\mu_k^{\dagger\#} = \xi_k/N$ for $\xi_k \sim \Pois(N\cdot \mu_k^\dagger)$ (\emph{not} i.i.d.). For finite $N$, each $\mu_k^{\dagger\#}$ is eventually zero for large $k$, and therefore $\mu_k^{\dagger\#} \bowtie 0$, implying convergence to the principal QSD $\mu^*$; and indeed, this is what we see in simulations: the time evolution $(M^*)^n[\mu^{\dagger\#}]$ initially tracks closely with the non-standard QSD $\mu^\dagger$, but after some random critical time $n$ eventually departs and instead converges quickly to the principal QSD $\mu^*$.
\end{itemize}
These figures are excluded mainly for brevity;
in each case the asymptotic behaviour is found to be consistent with the predictions laid out in Sec.~\ref{sec:quant}; namely, that a distribution is naturally inclined towards the QSD in its shape equivalence class. These QSDs -- even the non-principal ones -- are themselves, it seems, at least marginally stable to perturbation by common types of noise, although the flavour of noise is indeed an important factor.
For more discussion, see \cite{brevitt_weak_2026}.

\subsection{Convergence in the absence of a similarly-shaped QSD}

Finally, we want to investigate how a density with $\mu_k \bowtie \e^{-\sigma k}$ behaves when there is no QSD within its equivalence class. This is of particular interest as it includes the case of initialising from conditions $\mu_k \bowtie 2^{-k}$ or faster (which naturally occur from, e.g., initial fixed states $X_0=i$) in those regions of parameter space where no QSD of that shape exists (cf.\ Sec.~\ref{sec:bddns-qsd}): the red, green and white regions of Fig.~\ref{fig:map}.

\begin{figure}[ht]
    \centering
    \includegraphics[width=0.8\linewidth]{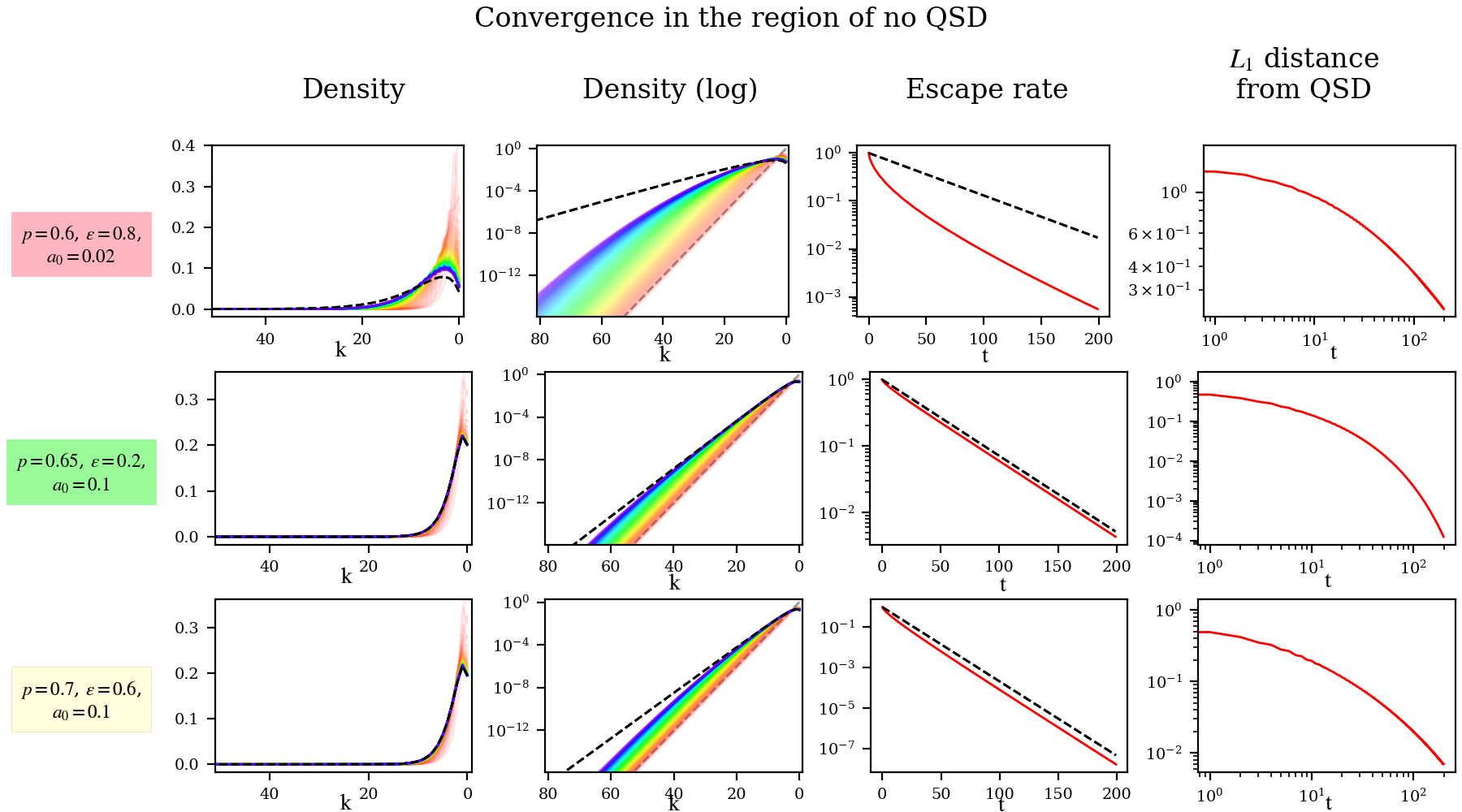}
    \caption{Convergence to the principal QSD $\mu^*$ from an initial density $\mu_k$ when $\mu_k\ll\mu^*_k$. In each row, $\mu_k \sim 2^{-k}$ while in each of these three regions $\mu^*_k\gg 2^{-k}$. In the first two columns, higher iterations are shown in higher frequency hues, see text for details.}
    \label{fig:test5}
\end{figure}

In Fig.~\ref{fig:test5} we initialise $\mu_k = 2^{-k-1}$ (we do not perturb this, as there is no need), and evolve it for $200$ timesteps. Increasing iterations $(M^*)^n[\mu]$ are plotted, on both a linear (first column) and a semi-log (second column) axis, in increasing frequency of hue, with the initial density in red and the latest iterations for large $n$ in blue. We do this for each of the three sampling points $(p,\ve)$
which do not support a QSD of this shape.
(The other two points, not shown, converge uninterestingly to the QSD within their class exponentially in time.)

In these cases we see a particular type of non-uniform convergence, in which the tail of $(M^*)^n[\mu]$ remains asymptotic to $\mu_k \sim 2^{-k}$, but the transient slowly converges to become closer to the principal QSD, $\mu^*$, which has the tail $\mu^*_k \gg 2^{-k}$. The escape rate of the system -- effectively the gradient of the line in Fig.~\ref{fig:test5} -- therefore quickly adopts the escape rate of $\mu^*$ (rather than that of $\mu$, since the escape rate is governed by the probability mass in and near $\mu_0$), while maintaining the asymptotic shape of $\mu$ for large $k$, and we find the $\ell^1$ distance $\norm{(M^*)^n[\mu]-\mu^*}_1$ decays sub-exponentially (perhaps as a power law) with respect to time, as opposed to the faster exponential rate we observe in the other parameter cases (not shown).

\section{Quasi-ergodic distributions (QEDs)} \label{sec:qed}


Finally, in this section we examine the existence of quasi-ergodic distributions (QEDs) in our system.
These measures
represent the distribution over phase space of the history of a surviving trajectory over its lifetime, see Sec.~\ref{sec:intro}. Unlike QSDs, a QED supported on all states of a Markov process, if it exists, should be unique,
for theoretical reasons.\footnote{
Because over a long lifetime the process must (a.s.) visit all states, the time average \eqref{eq:qed-def} if it exists is independent of its initial state in the long-time limit.
} A review of the existing literature for QEDs in countable substochastic chains is given in Sec.~\ref{sec:intro}, based on results in \cite{seneta_quasi-stationary_1966, vere-jones_ergodic_1967},
which give existence and uniqueness of a QED given by $\bm{q}_i=b_i\mu^*_i$, for $b_i$ the leading eigenvector of $M^T$, if $M$ is $R$-positive. In the language of dynamical systems, $M^T$ is sometimes called the \emph{Koopman operator} \cite{castro_existence_2024} in contrast to the \emph{Perron-Frobenius}, or transfer, operator $M$ \cite{collet_quasi-stationary_2013}.




The Koopman operator $M^T$ for our system is exactly the transpose of the transfer operator \eqref{eq:pelikan-markov}:
\begin{equation} \label{eq:koopman}
    M^T_{ji} = p\,\bm{1}_{j=i+1} + q\,\bm{1}_{j=i-1}
    + p(1-\ve)2^{-i-1}\,\bm{1}_{j=0}
\end{equation}
which may be written in generating function form, via $B(z):=\sum_{k=0}^\infty b_kz^k$, as
\begin{equation}
    [M_*(B)](z) = {p z B(z)} + {q \frac{B(z)-b_0}{z}}
    + {p \frac{(1-\ve)}{2} b_{1/2}}
\end{equation}
subject to $B(0)=b_0$ and $B(\frac{1}{2}) =: b_{1/2}$.
Let us neglect for now any constraints regarding physicality, such as positivity or normalisation.
Then we solve for the eigenvectors of $M_*$,
\begin{gather}
    \lambda B = M_*(B) = pzB + q\frac{B-b_0}{z} + p\frac{1-\ve}{2}b_{1/2} \nonumber \\
    \implies B(z) = \frac{p\frac{(1-\ve)}{2}zb_{1/2} - qb_0}{\lambda z - pz^2 - q}
\end{gather}
with $B(0) = (-qb_0)/(-q) = b_0$
as required, for $p<1$, and
\begin{equation}
    B(\tfrac12) = b_{1/2} = \frac{\tfrac14 p(1-\ve)b_{1/2}-qb_0}{\tfrac{\lambda}{2} -\tfrac{p}{4} -q} 
    \implies b_{1/2} = \frac{-qb_0}{\tfrac{\lambda}{2} -\tfrac{p}{2} -q +\tfrac{p\ve}{4}}
\end{equation}
which gives
\begin{equation}
    B(z) = \frac{\frac{-pqb_0(1-\ve)}{\lambda -p -2q +p\ve/2} z - qb_0}{\lambda z - pz^2 - q}.
\end{equation}
The normalisation of this measure has no physical meaning, since we will later multiply this measure by the QSD $\mu^*$ to obtain the QED $\bm{q}$, which will then be normalised. Hence we remove the constant prefactor $b_0$ and divide through by $q$ to obtain
\begin{equation}
    B(z) = \frac{
    1+\frac{p(1-\ve)}{\lambda -p -2q +p\ve/2} z
    }
    {1-\frac{\lambda}{q} z + \frac{p}{q}z^2}.
\end{equation}
However note that $\lambda$ remains a function of $b_0$, and remains a free variable just as $\mu_0$ in the case of Sec.~\ref{sec:2.1}.

\begin{figure}[ht]
    \centering
    \includegraphics[width=0.98\linewidth]{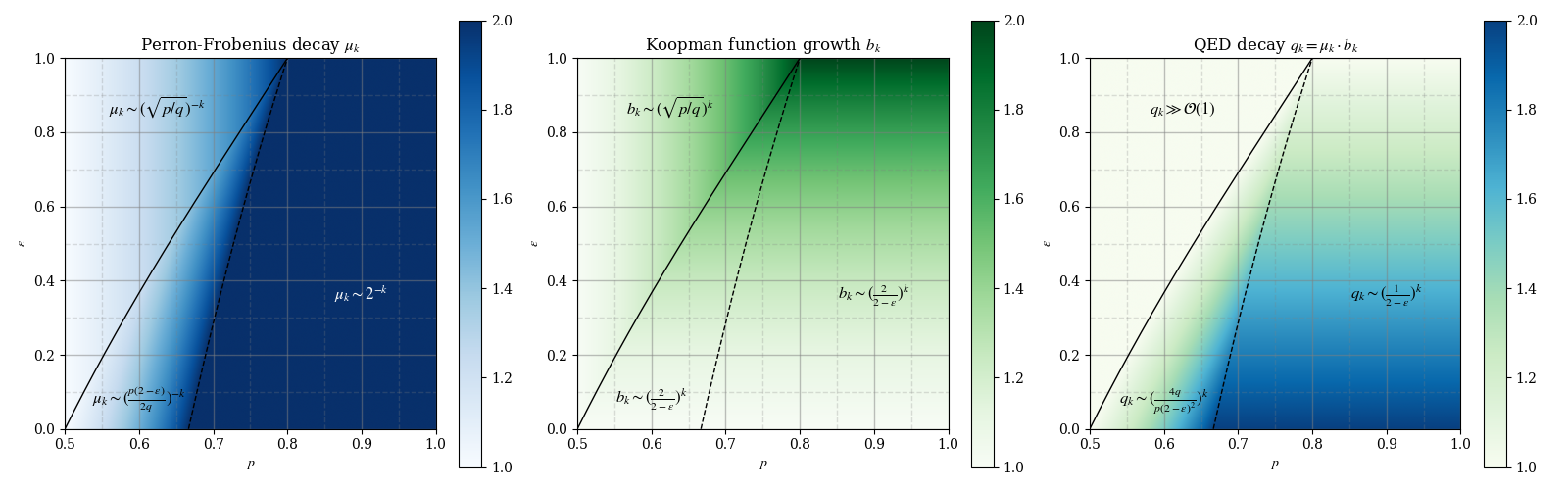}
    \caption{From left to right, shown as colour gradients as a function of $(p,\ve)$: decay rate (`shape') of the principal QSD $\mu^*_k$ wrt.\ $k$ (darker is faster decay); growth rate of the corresponding Koopman eigenvector $b_k$ wrt.\ $k$ (darker is faster growth); decay rate of the resulting QED $\bm{q}_k = \mu_k \cdot b_k$ (white corresponds to $\mu_k \cdot b_k \bowtie 1$; darker blue corresponds to faster rates of decay).}
    \label{fig:color-plot-6-parts}
\end{figure}

This function also will give us a variety of solutions depending on the eigenvalue $\lambda$. However, we focus on the eigenvalue $\lambda_u$ obtained from the principal escape rate, as calculated in Sec.~\ref{sec:bddns-qsd}; for technical reasons\footnote{namely, that the QED produced by multiplying \eqref{eq:koopman-exponentialregime} with its associated QSD is normalisable only because of the cancellation of terms $(z-\frac{2q}{p(2-\ve)})$ in \eqref{eq:koopman-exponentialregime}, which only occurs in the special case $\lambda=\lambda_u$. Without this, there is an additional pole in $B(z)$ which ensures the resulting QED is not normalisable; hence $\lambda=\lambda_u$ and \eqref{eq:koopman-exponentialregime} as written is the only suitable solution.}
we believe this is in fact the \emph{only} eigenvalue which produces a normalisable QED. This eigenvalue is given to be $\frac{4-4p\ve+p\ve^2}{4-2\ve}$ in the region of $R$-recurrence, and $\sqrt{4pq}$ in the region of $R$-transience. Hence, in the $R$-recurrent regime,
\begin{equation} \label{eq:koopman-exponentialregime}
    B(z) = \frac{
        1+\frac{p(1-\ve)(4-2\ve)}{(4-4p\ve+p\ve^2)+(-p-2q+p\ve/2)(4-2\ve)}z
    }{
        1-\frac{4-4p\ve+p\ve^2}{2q(2-\ve)}z +\frac{p}{q}z^2
    }
    = \frac{
        1+\frac{p(2-\ve)}{-2q}z
    }{
        \frac{p}{q} (z-\frac{2-\ve}{2}) (z-\frac{2q}{p(2-\ve)})
    }
    = \frac{q/p}{z-\frac{2-\ve}2};
\end{equation}
and in the $R$-transient regime,
\begin{equation} \label{eq:koopman-mixedregime}
    B(z) = \frac{1+\frac{p(1-\ve)}{\sqrt{4pq}-p-2q+p\ve/2}z}{1-\frac{\sqrt{4pq}}{q}z+\frac{p}{q}z^2}
    = \frac{1+\frac{p(1-\ve)}{\sqrt{4pq}-p-2q+p\ve/2}z}{\frac{p}{q}(z-\sqrt{q/p})^2},
\end{equation}
implying $b_k$ has shape given by exponential growth rates of $(\frac{2}{2-\ve})^k$ and $(\frac{1}{\sqrt{q/p}})^k$ respectively. Recall from \eqref{eq:pfdecay-1} and \eqref{eq:pfdecay-2} that the QSDs in these two cases were explicitly calculated as
\begin{equation}
    A(z) = \frac{(1-2p+\frac{p\ve}{2})}{q(2-z)(z-\frac{p(2-\ve)}{2q})}
    \quad \textrm{and} \quad
    A(z) = \frac{(1-\sqrt{4pq})(2-(2-\ve)z)}{\frac{\ve}{q}(2-z)(z- \sqrt{p/q} )^2}
\end{equation}
respectively, implying shapes of $\mu^*_k$ given by exponential decay rates of $2^{-k}$ or $(\frac{p(2-\ve)}{2q})^{-k}$ in the $R$-positive regime (delineated as described in Sec.~\ref{sec:bddns-qsd}) and $(\sqrt{p/q})^{-k}$ in the $R$-transient regime. Therefore, multiplying the Koopman eigenvector $b$ by the corresponding QSD $\mu^*$ gives a QED $\bm{q}$ which is positive and has a shape decaying as $\bm{q}_k\sim(\frac{1}{2-\ve})^k$ or $\bm{q}_k\sim(\frac{4q}{p(2-\ve)^2})^k$ in the $R$-recurrent regime, depending on the shape of $\mu^*$, and in the $R$-transient regime is not decaying exponentially at all, with $\bm{q}\bowtie 1$; looking at higher order terms, numerically, we see it is actually increasing subexponentially. In the $R$-transient regime, therefore, there is no normalisable QED. Observe that $(\frac{1}{2-\ve})^k$ is always (for $\ve<1$) exponentially decreasing,
as is $(\frac{4q}{p(2-\ve)^2})^k$ in the region where this rate takes effect,
and therefore the resulting QED in the $R$-recurrent ($R$-positive) regime is always normalisable.
These growth/decay rates are shown as colour gradients in Fig.~\ref{fig:color-plot-6-parts}.

\begin{figure}[ht]
    \centering
    \includegraphics[width=0.98\linewidth]{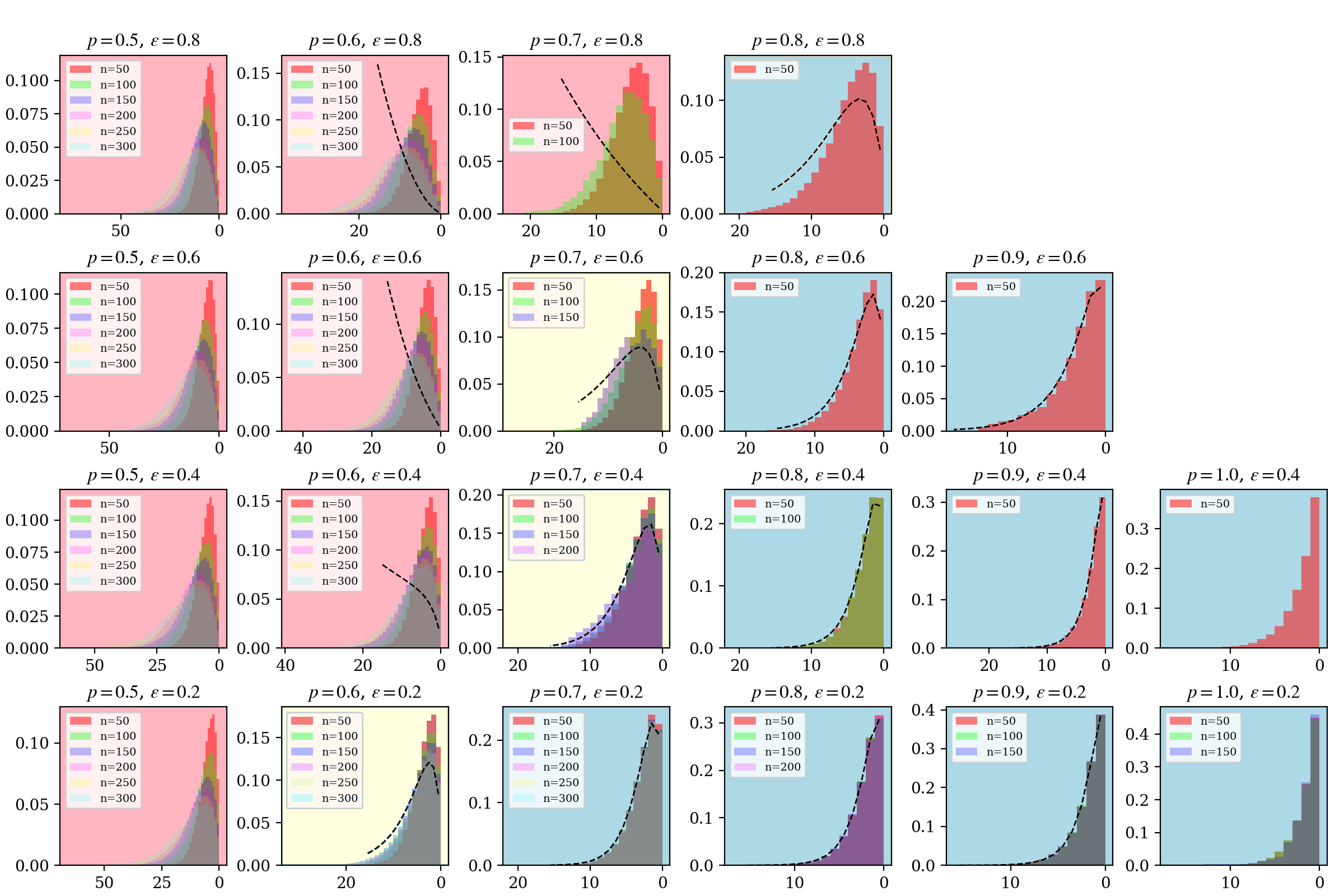}
    \caption{QEDs estimated from numerical simulations,
    for increasing measurement times up to between $t=50$ and $t=300$ (coloured histograms); compared against theoretical QEDs calculated analytically in Sec.~\ref{sec:qed} (black dashed lines). 
    Regions are colour coded according to $R$-recurrence and $R$-transience, and the shape of the principal QSD, see Fig.~\ref{fig:map} and Sec.~\ref{sec:bddns-qsd}.
    The calculations produce a normalisable QED exactly when the process is $R$-positive (blue and white regions).}
    \label{fig:ergodic1}
\end{figure}

Using a recurrence relation to accurately generate both $\mu^*$ and $b$, and their product $\bm{q}$, numerically, we produce the theoretical QEDs implied by the above equations, shown by black lines in Fig.~\ref{fig:ergodic1}. In the same figure, we test these results against simulations, which are generated by plotting histograms of the positions over time of an ensemble of trajectories, for each $(p,\ve)$, which survive up to time $t$, for varying $t$ which are plotted overlapping in different colours. Each subplot corresponds to a different $(p,\ve)$ and has its background colour coded based on the region, as seen in Fig.~\ref{fig:map}, into which it falls. In the region of $R$-recurrence, where there are normalisable QEDs (blue and white regions in Fig.~\ref{fig:ergodic1}), our calculated QEDs appear to match well against simulations,
in the sense that the simulated QEDs appear to converge with $t$ to a distribution well approximated by our calculations;
however, near the borderline, this convergence does appear to be slow.
Where the process is $R$-transient, and our calculated `QEDs' diverge (red region in Fig.~\ref{fig:ergodic1}), it appears that our simulations do not converge to any distribution at all for increased $t$, instead becoming increasingly flatter over time. In this case, we are therefore inclined to conclude that a QED does not exist in this dynamical regime,
in keeping with our intuition.

\section{Conclusion} \label{sec:c4concl}

In this paper we demonstrated the existence of a continuous spectrum of distinct QSDs for a relatively simple countable substochastic Markov chain. These multiple QSDs appear for all parameter values, regardless of $R$-recurrence or $R$-transience, and each is supported by its own eigenvalue, determining its own distinct rate of escape. The conditional invariance of each of these measures, and their stability to perturbation by random noise, is attested analytically and in simulations. We also analytically constructed unique QEDs for those parameters where they are known to exist -- namely, when the system is $R$-positive -- which are also attested in simulations.


It is worth considering the full implications of this result; namely, this means that \emph{in principle} the escape rate of the system, even in the long term, can be controlled, altered or prescribed without changing any parameters of the system, simply by varying the distribution of initial conditions. This also means that other quantities of a system, such as the Lyapunov exponent or metric entropy, may also be strongly dependent on the system's initial arrangement.
Our example, which is motivated by a simple spatially-continuous random map, the Pelikan map, should therefore be of some interest to researchers in a wide range of fields, from absorbing Markov chains and stochastic resetting to deterministic dynamical systems and statistical physics. Trivially, any random dynamical system can be expressed as a deterministic system of higher dimension, via a pseudo-skew product \cite{pelikan_invariant_1984}.
These results are also particularly significant due to the well-known intermittency produced by the Pelikan map. Whether the existence of multiple QSDs has a significant effect on the internal chaos dynamics of the system, if indeed such a question can be well-defined, is unclear to us at present, and would be a worthy subject for pursuit in future research. The internal dynamics of the random Pelikan system, in terms of Lyapunov exponents and recurrence times, was investigated extensively in \cite{brevitt_weak_2026}.


This result, although peculiar, is supported by literature, as outlined in Sec.~\ref{sec:intro}.
Nevertheless, we may remark that, in private correspondence with experts, the mere existence of multiple QSDs, all
supported across the entire phase space, has generally been met with a degree of surprise and even scepticism, particularly within the (random) dynamical systems community.
To our knowledge, this is the first explicit and concrete example in the literature of such phenomena naturally occurring in a non-trivial dynamical system.


\ack{The authors would like to acknowledge very helpful private discussions with Martin Rasmussen (Imperial College London) and Oscar Bandtlow (Queen Mary University of London) which contributed to this research.}




\addcontentsline{toc}{chapter}{\protect\numberline{}Bibliography} 
\printbibliography[title={References}]

\end{document}